\documentclass[11pt,dvipsnames]{article}
\usepackage{verbatim,amsmath,amssymb}
\usepackage{epsfig,float,color}
\usepackage{geometry}
\usepackage{setspace}
\usepackage{natbib}
\usepackage[title,titletoc]{appendix}
\usepackage{breakcites}
\usepackage{amsthm,mathrsfs,amsfonts,dsfont}
\usepackage[yyyymmdd,hhmmss]{datetime}
\usepackage[bookmarks=true]{hyperref}
\usepackage{bookmark}
\usepackage{cleveref}
\usepackage[font=small,labelfont=bf]{caption}
\usepackage{subcaption}
\usepackage{tablefootnote}
\usepackage{amssymb}
\usepackage{pifont}
\usepackage{nicematrix}
\usepackage{booktabs}
\usepackage{array}
\usepackage{float}
\usepackage[section]{placeins}
\usepackage{mmap}
\theoremstyle{remark}

\definecolor{darkblue}{rgb}{0,0,1}
\hypersetup{colorlinks=true, breaklinks=true, linkcolor=darkblue, menucolor=darkblue, citecolor=darkblue, urlcolor=darkblue}

\newcommand{\bitm}{\begin{itemize}}
\newcommand{\eitm}{\end{itemize}}
\newcommand{\bnumr}{\begin{enumerate}}
\newcommand{\enumr}{\end{enumerate}}

\newcommand {\eqb}[1]{\begin{equation}\begin{array}{#1}}
\newcommand {\eqe}{\end{array}\end{equation}}

\newcommand {\esb}[1]{\begin{equation*}\begin{array}{#1}}
\newcommand {\ese}{\end{array}\end{equation*}}
\newcommand {\ds}{\displaystyle}

\newcommand {\dif}{\mathrm{d}}

\newcommand {\II}{{I\kern-.3em I}}
\newcommand {\III}{{I\kern-.3em I\kern-.3em I}}

\newcommand {\mra}{\mathrm{a}}
\newcommand {\mrb}{\mathrm{b}}
\newcommand {\mrc}{\mathrm{c}}

\newcommand {\IR}{{\rm\kern.24em
   \vrule width.02em height1.53ex depth-.05ex
   \kern-.3em R}}
\newcommand {\ic}{{\rm\kern.20em
   \vrule width.02em height1.0ex depth-.05ex
   \kern-.22em c}}
\newcommand {\ia}{{\rm\kern.20em
   \vrule width.02em height1.05ex depth-.0ex
   \kern-.25em a}}
\newcommand {\IC}{{\rm\kern.24em
   \vrule width.02em height1.4ex depth-.05ex
   \kern-.26em C}}
\newcommand {\ID}{{\rm\kern.34em
   \vrule width.02em height1.5ex depth-.05ex
   \kern-.36em D}}
\newcommand {\IS}{{\rm\kern.24em
   \vrule width.02em height1.6ex depth.05ex
   \kern-.26em S}}
\newcommand {\IT}{{\rm\kern.50em
   \vrule width.02em height1.55ex depth-.05ex
   \kern-.52em T}}

\newcommand {\IE}{{\rm\kern.24em
   \vrule width.02em height1.55ex depth-.05ex
   \kern-.33em E}}
\newcommand {\IEa}{{\rm\kern.24em
   \vrule width.02em height1.55ex depth-.05ex
   \kern-.33em E}^{1}_{ijkl}}
\newcommand {\IEb}{{\rm\kern.24em
   \vrule width.02em height1.55ex depth-.05ex
   \kern-.33em E}^{2}_{ijkl}}

\newcommand {\Ass}[2]{\kern 0.9ex \vrule width0.45em height0.2ex depth0ex \kern -2.1ex \bigwedge_{#1}^{#2}}
\newcommand {\ASS}[2]{\kern 1.45ex \vrule width0.5em height0.2ex depth0ex \kern -2.65ex \bigwedge_{#1}^{#2}}

\graphicspath{ {images/} }

\begin{document}

\begin{center}
\Large{\bf{A new efficient ab-initio approach for calculating the bending stiffness of 2D materials}}\\

\end{center}

\renewcommand{\thefootnote}{\fnsymbol{footnote}}

\begin{center}
\large{Farzad Shirazian$^\mra$ and Roger A.~Sauer$^{\mra,\mrb,\mrc,}$\footnote[1]{corresponding author, email: roger.sauer@pg.edu.pl, sauer@aices.rwth-aachen.de}}\\
\vspace{4mm}

\small{\textit{
$^\mra$Aachen Institute for Advanced Study in Computational Engineering Science (AICES), \\ 
RWTH Aachen University, Templergraben 55, 52056 Aachen, Germany \\[1.1mm]
$^\mrb$Faculty of Civil and Environmental Engineering, Gda\'{n}sk University of Technology,\\ ul.~Narutowicza 11/12, 80-233 Gda\'{n}sk, Poland \\[1.1mm]
$^\mrc$Dept.~of Mechanical Engineering, Indian Institute of Technology Guwahati, Assam 781039, India
}}

\end{center}

\vspace{-4mm}

\renewcommand{\thefootnote}{\arabic{footnote}}

\vspace{3mm}

\vspace{3mm}
\rule{\linewidth}{.15mm}

\textbf{Abstract}: This work proposes a new efficient approach for calculating the bending stiffness of two-dimen\-sional materials using simple atomistic tests on small periodic unit cells. The tests are designed such that bending deformations are dominating and membrane deformations are minimized. Atomistic ab-initio simulations then allow for the efficient computation of bending energies. Density functional theory is used for this. Atomistic bending energies are then compared to classical models from structural mechanics. Two different models are considered for this -- one based on beam theory and one based on rigid linkage theory -- and their results are compared with each other. Four different materials with 2D hexagonal (honeycomb) structure are chosen as a case study: graphene, hexagonal boron nitride, silicene, and blue phosphorene. The calculated bending stiffnesses converge with increasing unit cell size, such that small unit cells already provide accurate results that are in good agreement with the literature. Using the same atomistic tests, it is shown that the bending stiffness of graphene can still be considered constant at moderately large deformations. Apart from being efficient and accurate, the proposed approach allows for various extensions. 

\vspace{-4mm}
\rule{\linewidth}{.15mm}

\section{Introduction}
The discovery of graphene and its successful synthesis \citep{Novoselov2005_01} has led researchers to embark on extensive research on two-dimensional materials. These materials exhibit unique properties such as high carrier mobility \citep{Mir2020_01}, piezoelectricity \citep{Hinchet2018_01,Mortazavi2021_01}, tunable plasmons \citep{Grigorenko2012_01, GarciadeAbajo2014_01}, quantum Hall effect \citep{Goerbig2011_01}, and negative Poisson ratio \citep{Wang2017_02}. Depending on test conditions, the Poisson ratio of graphene has been reported positive \citep{Gui2008_01} and negative \citep{Jiang2016_01, Burmistrov2018_01, Jin2020_01}, which can be attributed to wrinkling caused by defects \citep{Grima2015_01} or stresses \citep{Burmistrov2018_01}. Over the past decade, many more two-dimensional materials with interesting properties have been discovered and studied \citep{Wang2012_01, Jiang2014_01, Akinwande2017_01, Zhang2017_01, Yang2017_01, Wang2017_01, Sorokin2021_01, Tong2021_01}. A criterion for the existence and stability of 2D materials is given in \citet{Chen2017_01}.\\
Theoretical and computational continuum models developed for these materials play an essential role in the study and design of nanostructures since they allow to study structures at larger time and length scales than what is practical using atomistic methods. Continuum models require accurate material parameters that need to be determined from ab-initio simulations \citep{Arroyo2004_01, Kumar2015_01, Ghaffari2018_01, Ghaffari2019_01,Shirazian2018_01}. With proper calibration, continuum models allow to describe graphene across a large range of deformations, as a recent comparison has shown \citep{Mokhalingam2020_01}.\\
Curved two-dimensional materials are especially of interest due to their structural stiffness \citep{Pini2016_01} and tunability \citep{Yu2016_01}, and they serve as a building block for nanostructures such as nanotubes and nanocones. Two-dimensional materials exhibit curvature-dependent bending Poisson effect \citep{Liu2014_01}, and even a negative bending Poisson ratio has been reported for C3N \citep{Chen2020_01}. Bending stiffness is an essential parameter to describe the behavior of 2D materials. 
Two atomistic approaches for the calculation of the bending stiffness of 2D materials are commonly used: (i) computing the bending energy of nanotubes of different radii \citep{Munoz2010_01} and (ii) applying curvature or indentation increments to a large surface with free edges \citep{Scarpa2010_01}. In the first case, simulations can be very expensive, especially for the study of small curvatures, since more atoms are required to generate nanotubes with small curvatures (i.e.~large radii). In the second case, a large supercell has to be used, and the effects of free boundaries need to be dealt with. As a result, both approaches are not ideal for accurate but expensive methods such as density functional theory (DFT), whose simulation cost scales as $O(N^3)$ with the number of atoms \citep{sholl2011_01}.\\
Therefore, a new approach to this problem is proposed here that applies small curvatures to small unit cells with periodic boundary conditions. The approach is efficient, free of boundary effects, characterized by smooth deformations, and applies to all 2D lattice materials. One of the challenges of developing continuum models from atomistic data is establishing a link between the continuous nature of the continuum model and the discreteness of the atomistic unit cell. In order to address this, two different structural models for the analysis of the atomistic tests are suggested and compared. In order to validate each model, the consistency and convergence of the results among different atomic configurations are assessed.\\
The proposed framework can be used for the calculation of the bending modulus of two-dimensional materials based on atomistic ab-initio simulations that 	
\begin{itemize}
	\item require few atoms, making ab-initio simulations efficient;
	\item are periodic along the in-plane directions, which eliminates free boundary effects;
	\item are dominated by bending, i.e., membrane energies are negligible; and 
	\item are suitable for the study of a large range of curvatures.		
\end{itemize}
The resulting continuum bending models can then be used to study 2D materials under arbitrary and non-uniform bending conditions.\\
The remainder of this section presents a review of existing research related to this work. The design of atomistic tests and the details of DFT simulations are elucidated in Secs.~\ref{s:A_b_t} and \ref{s:Atomistic_Simulations}, respectively. In order to develop the continuum bending models, two approaches, a beam (Sec.~\ref{s:Beam_Model}) and a rigid linkage (Sec.~\ref{s:RL_Model}) model, are presented for the bending analysis and calibration of its model parameters. In Sec.~\ref{s:Results}, the presented approach is then used for the determination of bending stiffnesses for graphene (at small and large deformation), hexagonal boron nitride (h-BN), silicene, and blue phosphorene ($\beta$-P). This is followed by a detailed discussion in Sec.~\ref{s:Discussion} for each considered case. The paper concludes with Sec.~\ref{s:Conclusion}.

\subsection{Related works}
\citet{Berinskii2014_01} suggest modeling all the carbon bonds using elastic rods. The bending stiffness of graphene is then analytically estimated from the stiffness of the rods. However, Bernoulli-Euler beam theory in this framework does not lead to satisfactory results for graphene, and Timoshenko beam theory requires the determination of too many parameters. Unlike our approach, they consider cross sections, and effectively a thickness for the rods, and determine the properties of the rods from graphene's elastic parameters instead of atomistic bending tests. In \citet{Zhang2011_01} and \citet{Nikiforov2014_01}, the authors argue that the bending stiffness of graphene cannot be plausibly defined from 3D elasticity theory, but should be attributed to the torsional misalignment of the $\pi$ hybrid orbitals. This implies that continuum methods are not sufficient for determining the bending stiffness of 2D materials.\\
An alternative to continnum methods are atomistic methods based on interatomic potentials. \citet{Arroyo2004_01} calculate the bending modulus of graphene from bond-order inter-atomic potentials while characterizing the deformation using the exponential Cauchy-Born rule. \citet{Zhang2015_01} analytically derive the elastic bending stiffness of single-layer black phosphorene from the atomic interactions described by the valence force field model. \citet{Davini2017_01} determine an equivalent plate equation for the deformations of graphene and use the 2nd-generation Brenner potential to calculate the bending stiffness and the Gaussian stiffness of graphene. As a result, an analytical expression of the Gaussian stiffness is provided. The thermomechanics of monolayer graphene and its out-of-plane thermal fluctuations are studied in \citet{Gao2014_01}, and the results from harmonic statistical mechanics are compared with NPT and NVT molecular dynamics simulations. The comparison shows the importance of anharmonic effects, especially at higher temperatures and for larger membranes. Considering these effects, \citet{Ahmadpoor2017_01} study the influence of thermal fluctuations on the bending stiffness and suggest a size-dependent effective bending stiffness for finite temperature using variational perturbation theory. \\
Although interatomic potentials can provide good results, they are still only approximations of the complex quantuum mechanical interactions and can become inaccurate, especially if they are used outside the range they have been designed for. 
To avoid such inaccuracies, ab-initio atomistic methods, such as DFT, should be used. \citet{Wei2013_01} use DFT simulations to calculate the bending stiffness of graphene from the energy of single-walled carbon nanotubes of different radii and use this value along with the energies of fullerenes of different sizes to determine the Gaussian bending stiffness of graphene. \citet{Koskinen2010_01, Koskinen2010_02} suggest to approximate the bending stiffness of two-dimensional materials by extending Bloch's theorem for curved surfaces, which is very effective for different types of low curvature distortions. Implementing their so-called revised periodic boundary conditions is, however, not straightforward in plane-wave DFT codes \citep{Kit2011_01} and density functional tight binding or classical methods need to be used instead. \citet{Banerjee2016_01} introduce a new method called cyclic density functional theory and develop a symmetry-adapted finite-difference method, which enables the study of cyclic symmetries using Kohn-Sham density functional theory. The application of this method to group-IV nanotubes and forty-four atomic monolayers in the low-curvature limit is presented in \citet{Ghosh2019_01} and \citet{Kumar2020_01}, respectively. These approaches apply the curvature to the unit cell, and, in contrast to our work, the curvature is not directly applied to the atomic configurations.\\
In addition to the mentioned approaches, the bending stiffness of two-dimensional materials can be determined experimentally through electrostatic actuation \citep{Lindahl2012_01}, folding tests \citep{Zhao2015_01}, suspended indentation tests \citep{Chandler2020_01}, or deformation over a step \citep{Han2020_01}, to name a few methods. 

\section{Methods}\label{s:Theory_Methods}
\subsection{Atomistic bending tests}\label{s:A_b_t}
\begin{figure}[H] 
	\centering
	\includegraphics[width=\textwidth]{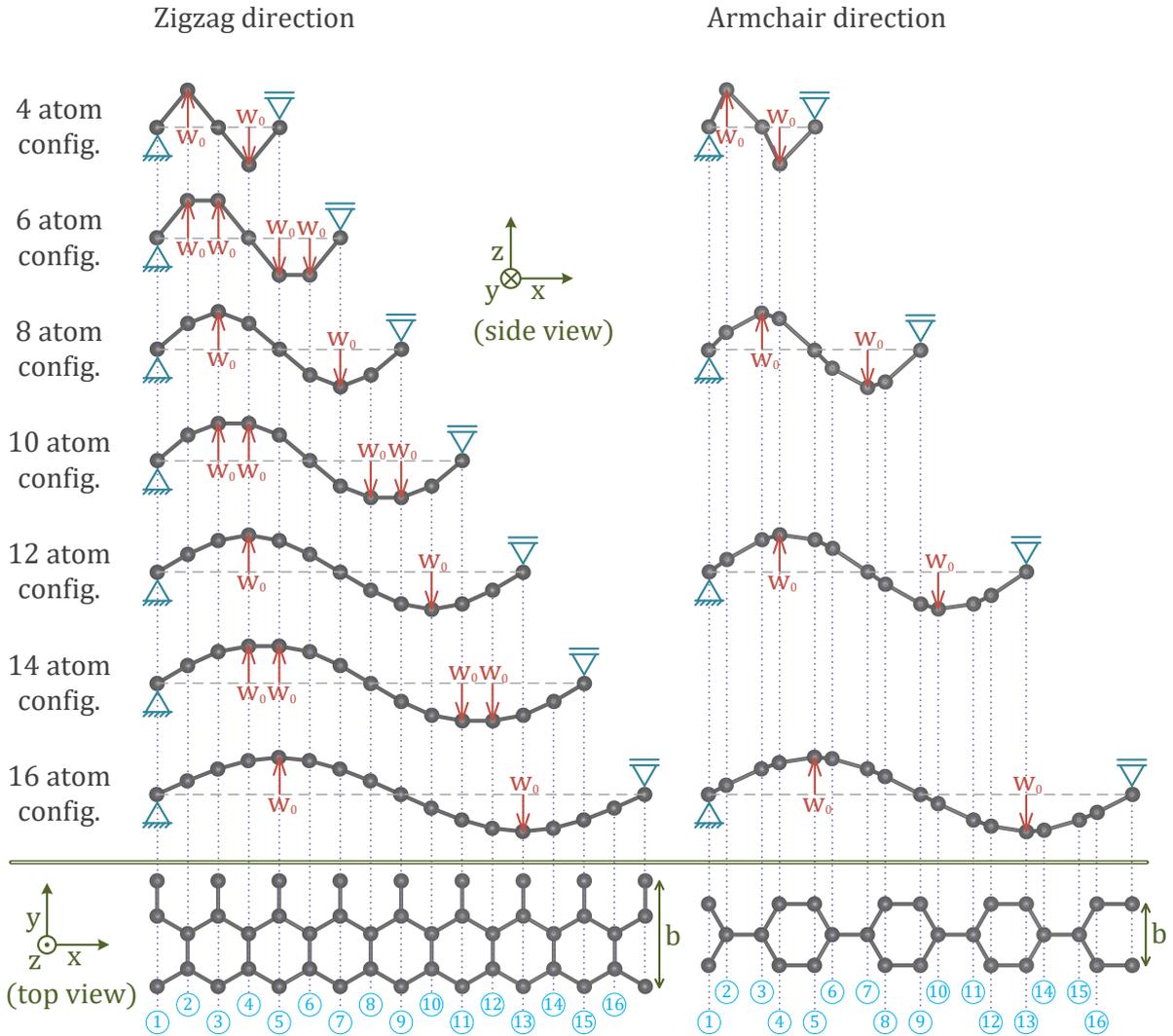}
	\vspace{- 6.5mm}
	\caption{Atomistic bending tests: atomic deflections and boundary conditions for 11 different atomistic test configurations. A predefined out-of-plane displacement $w_0$ is applied to certain atomic sites, as shown. The boundary atoms are fixed as shown. The rest of the atoms are allowed to relax. The configuration number $N$ is defined as the number of distinct atomic coordinates (visible atoms in the side view), along the $x$-direction in each unit cell. Therefore, $N = \frac{1}{2} \times$(number of atoms) for the zigzag and $N= $ (number of atoms) for the armchair direction. The atomic sites are numbered in blue.} \label{f:AllModesBC}
\end{figure}

In order to determine the bending stiffness from ab-initio data, atomistic tests are designed that cause uniaxial bending by applying the displacement $w_0$ to selected atoms, see Fig.~\ref{f:AllModesBC}. The test configurations are periodic in all directions, and they are designed such that the curvature can be arbitrarily low. As the figure shows, different atomic configurations can be considered: here, all possible periodic atomic configurations with up to 16 distinct atomic positions along the bending direction, which corresponds to the $x$-direction in Fig.~\ref{f:AllModesBC}. Considering different configurations allows for validating results by comparison. Ideally, the bending stiffness should be independent of the atomic configuration.\\
It can be seen that the largest unit cell in our proposed framework has 32 atoms, which is still much smaller than the number of atoms required for the classical approach based on nanotubes, especially at small curvatures: for example, the number of atoms for a nanotube unit cell with a curvature of 0.01 $\mathrm{nm}^{-1}$ is two orders of magnitude larger than in our proposed approach, which means the quantum mechanical computations for the classical approach would be up to six orders of magnitude larger than for our approach, as the scaling is $O(N^3)$ \citep{sholl2011_01}.

\subsection{Atomistic simulations}\label{s:Atomistic_Simulations}
\begin{figure}[h]
	\centering
	\begin{subfigure}[t]{0.28\textwidth}
		\centering 
		\includegraphics[width=\textwidth]{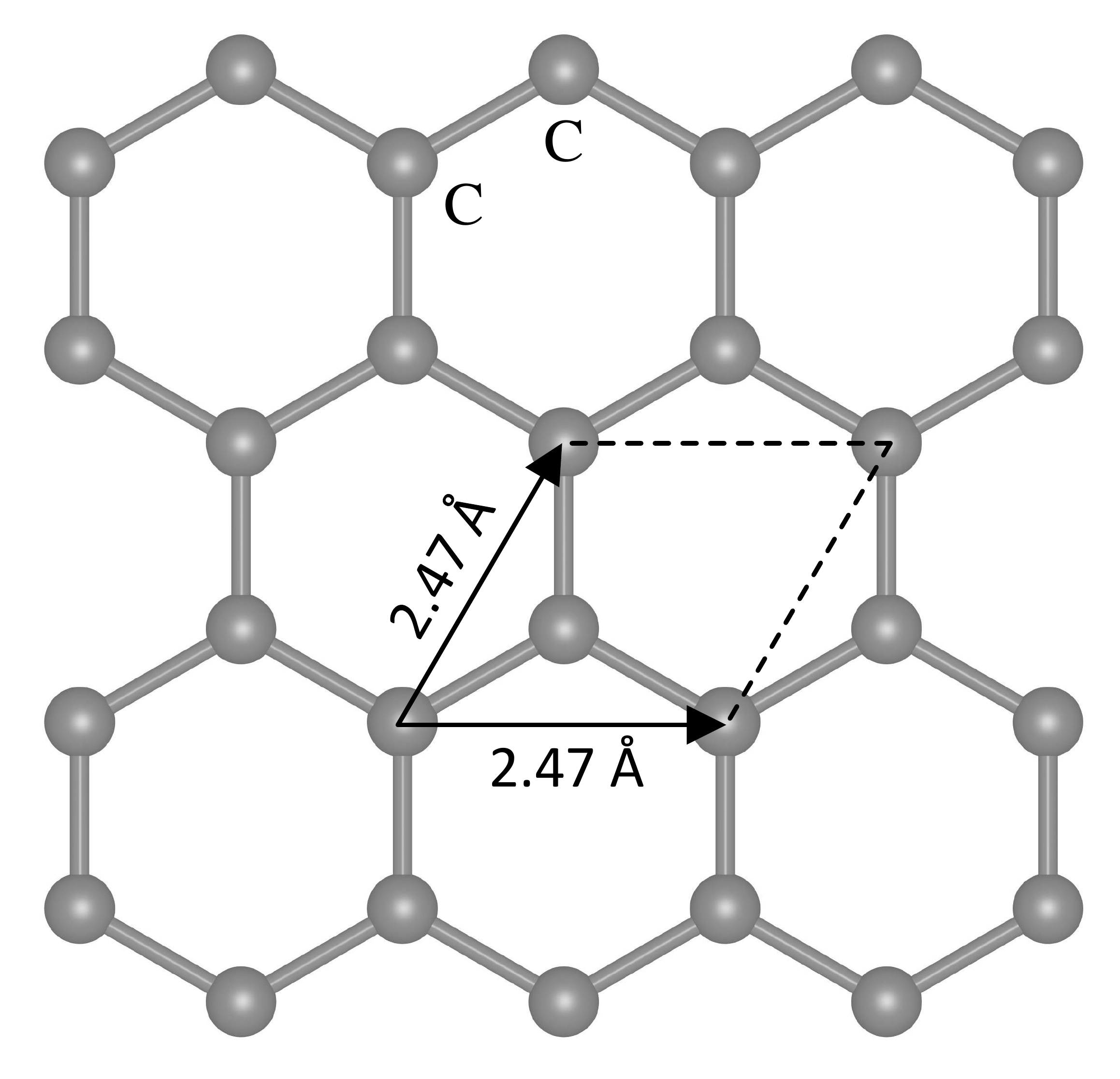}
		\vspace*{-6.5mm}
		\caption{Graphene}
		\label{f:Graphene}
	\end{subfigure}
	\hspace*{0.05\textwidth}
	\begin{subfigure}[t]{0.28\textwidth}
		\centering
		\includegraphics[width=\textwidth]{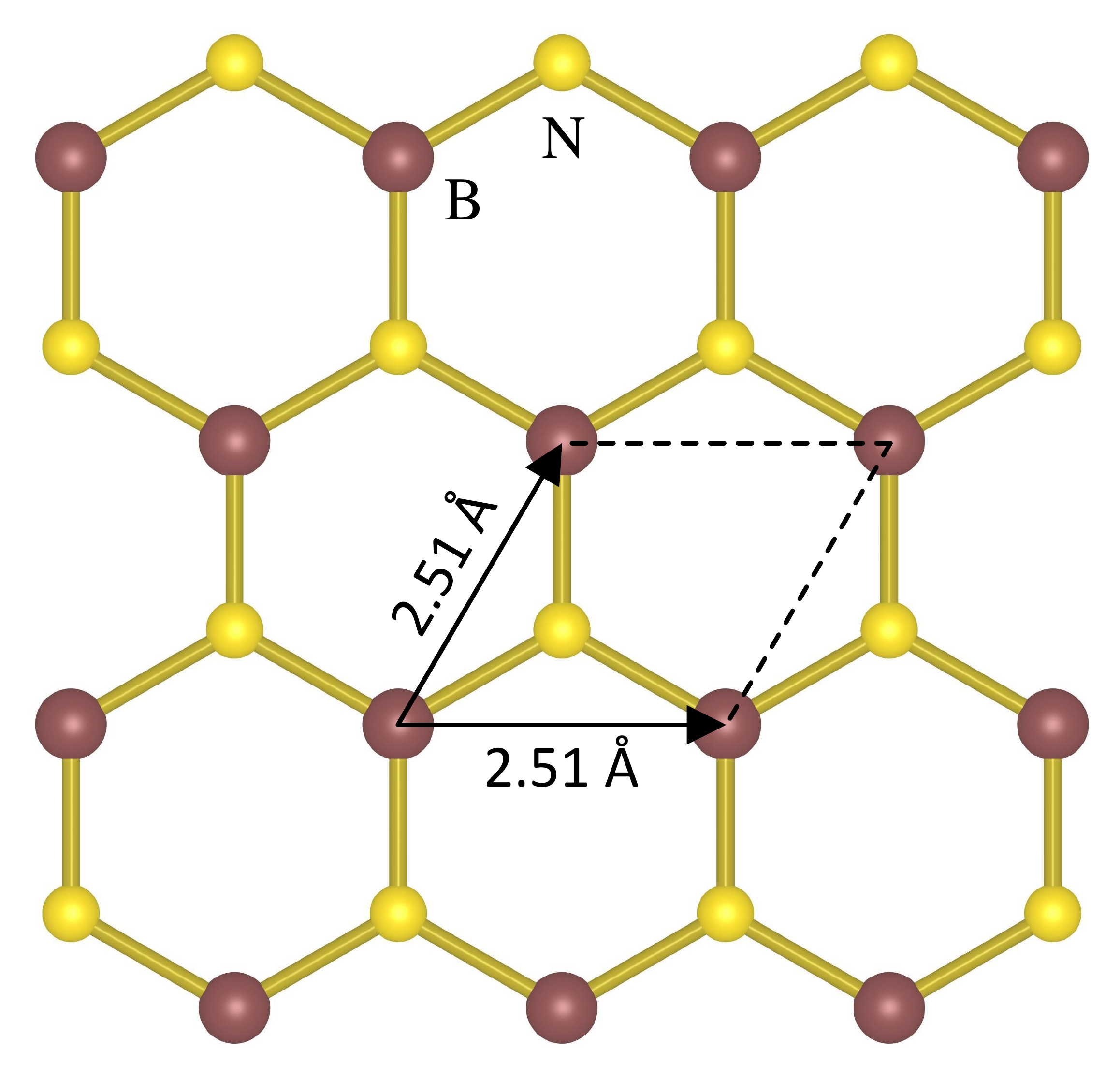}
		\vspace*{-6.5mm}
		\caption{h-BN}
		\label{f:hBN}
	\end{subfigure}
	\\
	\begin{subfigure}[t]{0.32\textwidth}
		\centering
		\includegraphics[width=\textwidth]{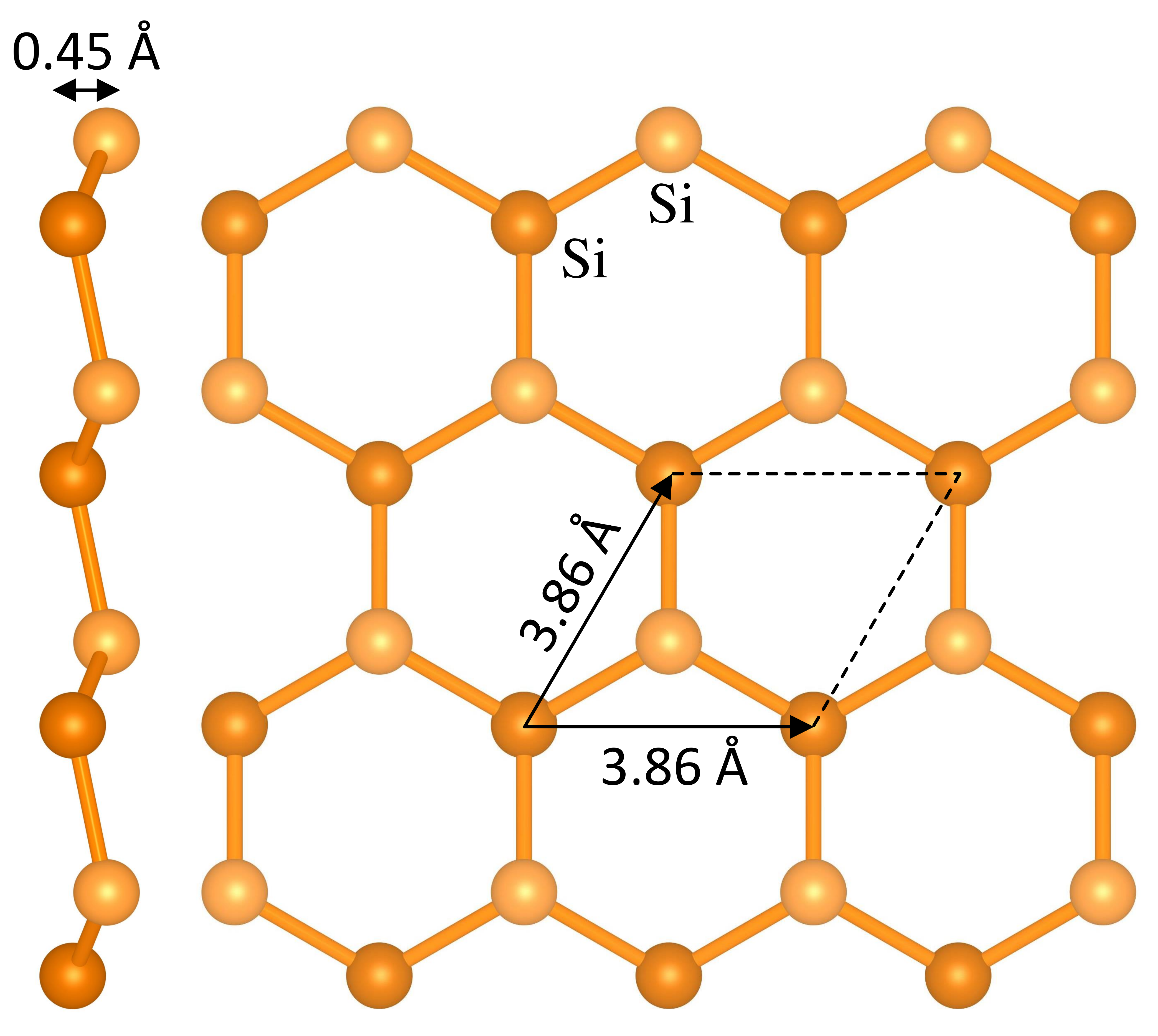}
		\vspace*{-6.5mm}
		\caption{Silicene}
		\label{f:Silicene}
	\end{subfigure}
	\hspace*{0.02\textwidth}
	\begin{subfigure}[t]{0.32\textwidth}
		\centering
		\includegraphics[width=\textwidth]{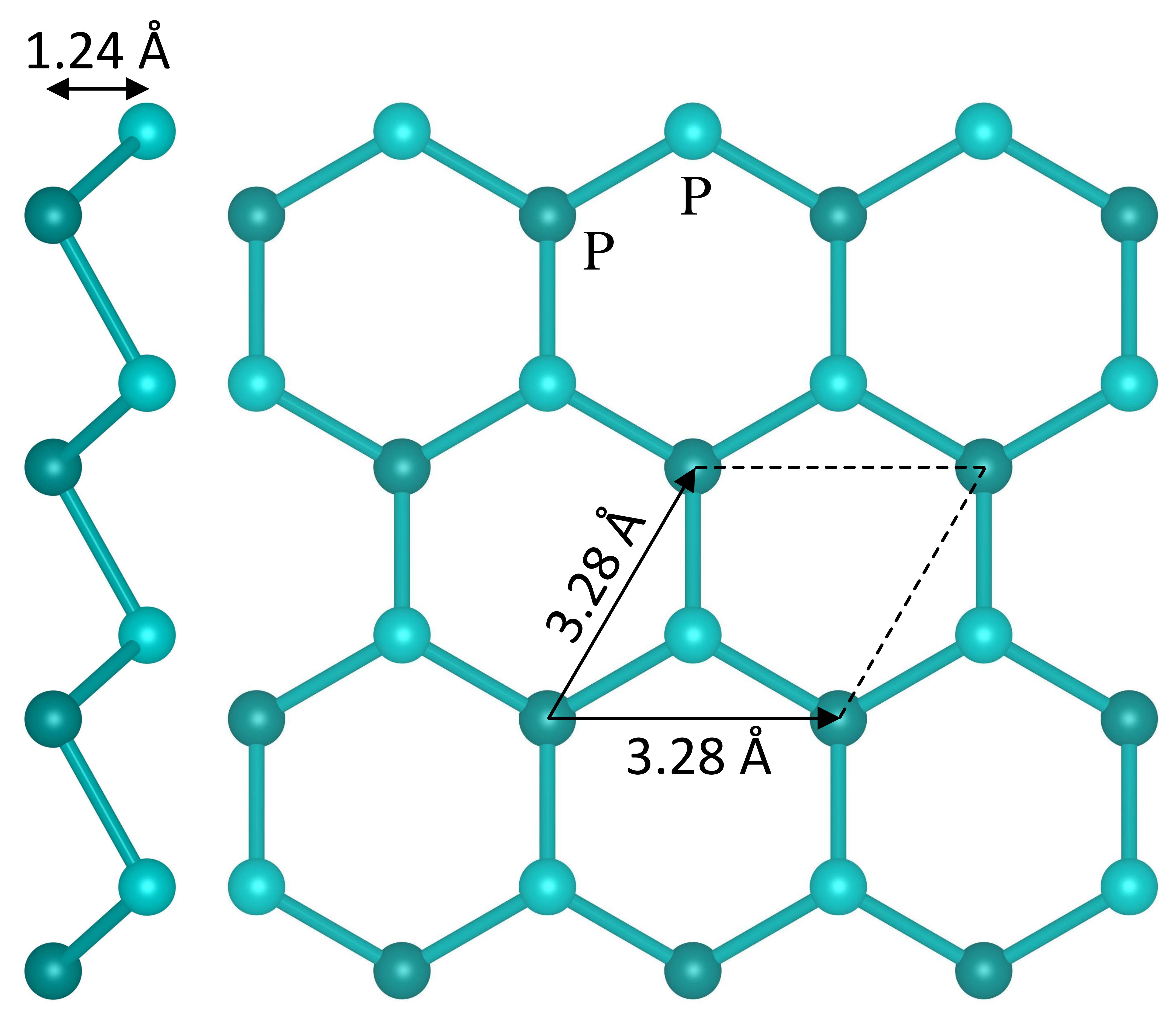}
		\vspace*{-6.5mm}
		\caption{Blue phosphorene}
		\label{f:BlueP}
	\end{subfigure}
	\vspace{0 mm}
	\captionsetup{width=.70\linewidth}
	\caption{Atomic structure of graphene, hexagonal boron nitride, silicene, and blue phosphorene. The lattice parameters are shown in the figures. Graphene and h-BN are flat, whereas silicene and blue phosphorene have a thickness of 0.45~\AA\ and 1.24~\AA, respectively. All atomic structures are visualized with VESTA \citep{Momma2011_01}.}  \label{f:hBNSilicene}
\end{figure}

The atomic configurations shown in Fig.~\ref{f:AllModesBC} are analyzed with DFT simulations at zero Kelvin using periodic unit cells. Displacement $w_{0}$ is increased in small increments to reach its final value. At each step, the unit cell dimensions and positions of free atoms are relaxed. The total energy of the unit cell is then calculated from DFT for the relaxed structure. \\
The DFT simulations are carried out using the {\sc Quantum Espresso} package \citep{Giannozzi2009_01,Giannozzi2017_01}. The Perdew-Burke\--Ernzerhof (PBE) exchange-correlation functional \citep{Perdew1996_01} is used to approximate the energy, and the effects of non-valence electrons are approximated using the Projected Augmented Wave (PAW) method \citep{Blochl1994_01}. The Brillouin zone integrations are performed using meshes generated by the Monkhorst-Pack scheme \citep{Monkhorst1976_01}. The k-point mesh for the primitive cell of the considered 2D materials and kinetic energy cutoff for the wave functions, respectively, are: $40 \times 40 \times 1$ and 60~Ry ($\mathrm{Ry}=13.606~\mathrm{eV}$) for graphene, $30 \times 30 \times 1$ and 100~Ry for h-BN, $40 \times 40 \times 1$ and 100~Ry for silicene, $40 \times 40 \times 1$ and 80~Ry for blue phosphorene. The k-point meshes are then adjusted for the unit cells of different atomic configurations in Fig.~\ref{f:AllModesBC}. The convergence threshold for the self-consistent field iteration is $10^{-8}$~Ry. The Coulomb interaction between the periodic replicas in the out-of-plane direction is truncated using the method by \citet{Sohier2017_01} and allowing a 30-35~$r_\mathrm{Bohr}$ ($r_\textrm{Bohr} = 0.5292~\mathrm{\AA}$) vacuum space between the periodic replicas.\\
The energy, force, and pressure convergence threshold for structural optimization are set to $10^{-8}$~Ry, $10^{-7}$~$\mathrm{Ry}/r_\mathrm{Bohr}$, and $10$~mbar, respectively. The initially undeformed and relaxed configurations of the studied materials are shown in Fig.~\ref{f:hBNSilicene}.

\subsection{Structural mechanics: Beam model} \label{s:Beam_Model}
The first method to calculate the bending stiffness uses beam theory. According to Euler-Bernoulli beam theory for small deformations, the out-of-plane deflection of a beam under uniaxial bending with boundary conditions as shown in Fig.~\ref{f:AllModesBC} is a piece-wise 3rd order polynomial along $x$ that can be written as
\eqb{lll} \label{e:zfx}
	\ds w=w(x) = a_3 x^3 + a_2 x^2 + a_1 x + a_0\,,
\eqe
where $a_i$ are constants. The curvature along the bending direction is given as
\eqb{lll} \label{e:kappa1Dapprox}
\ds \kappa = \frac{\dif^2 w}{\dif x^2}\\
\eqe
for small deformations.
Due to symmetry, it is sufficient to determine this polynomial in the range $0 \leq x \leq L$, where $4L$ is the period, i.e., the length of the unit cell. It is possible to decrease the number of unknowns ($a_i$) from four to one using the following three boundary conditions: 
\begin{itemize}
	\item $w=0$ at $x=0$,
	\item $\dif w / \dif x=0$ at $x=L$  to satisfy symmetry and $C^{1}$-continuity of the deflection curve,
	\item $M = 0$ at $x=0$, which implies $\kappa= 0$ according to the constitutive relation $M \sim \kappa$.
\end{itemize}

With this, the deflection and curvature of Eqs.~(\ref{e:zfx}) and (\ref{e:kappa1Dapprox}) simplify to
\eqb{lll}\label{e:BeamDeformation}
\ds w(x)=\alpha\bigg(3\,\frac{x}{L}-\Big(\frac{x}{L}\Big)^3\bigg)\\[3mm]
\eqe
and
\eqb{lll}
\ds \kappa(x) = - \frac{6 \alpha }{L^3}x\,,  \\
\label{e:PolyDeformation3}
\eqe
where $\alpha := - a_3 L^3$ is proportional to the applied displacement $w_0$. $\alpha$ follows from $w(x^*)=w_0$, where $x^*$ denotes the location of the applied displacement $w_0$, listed in Tab.~\ref{t:alfas}. Alternatively, $\alpha$ can be determined from minimizing the difference between all atomic positions obtained from DFT and Eq.~(\ref{e:BeamDeformation}). However, the difference between these alternatives was found to be negligible, which implies that the DFT deflection at the lattice position accurately follows the one predicted by beam theory.
\begin{table}[h] 
	\centering
	\small
	\caption{Values of $\hat{x} := x^*/L$, determining the location where $w_0$ is applied, and $\xi$, determining the energy in the rigid linkage model, for the 11 considered atomic configurations of Fig.~\ref{f:AllModesBC}. The beam deflection follows from Eq.~(\ref{e:BeamDeformation}) with $\alpha = w_0/(3\hat{x}-\hat{x}^3)$. Note that the values of $\hat{x}$ and $\xi$ depend on the atomic lattice structure, and the listed values correspond to 2D materials with hexagonal (honeycomb) structures.} \label{t:alfas} 
	\vspace{2mm}
	\begin{tabular}{lcccc}
		\hline \\ [-3mm]
		 & \multicolumn{2}{c}{Zigzag direction} & \multicolumn{2}{c}{Armchair direction}\\ [1mm]
		 \cmidrule(lr){2-3} 
		 \cmidrule(lr){4-5} 
		 &  $\hat{x}$ & $\xi$  & $\hat{x}$ & $\xi$ \\ [1mm]   
		\hline \\ [-3mm]
		4-atom-config.  & 1 & 4 &  2/3 & 5.6250  \\ [1mm]
		\hline \\ [-3mm]
		6-atom-config.  & 2/3 & 4.5 & & \\ [1mm] 
		\hline \\ [-3mm]
		8-atom-config.  & 1 & 8/3 &  1 & 2.9460 \\ [1mm]
		\hline \\ [-3mm]
		10-atom-config.  & 4/5 & 2.5 & & \\ [1mm]
		\hline \\ [-3mm]
		12-atom-config.  & 1 & 1.8945  &  8/9 & 1.9191 \\ [1mm]
		\hline \\ [-3mm]
		14-atom-config.  & 6/7 & 1.7505 & & \\ [1mm]
		\hline \\ [-3mm]
		16-atom-config.  & 1 & 1.4545 & 1 & 1.4400 \\ [1mm]
		\hline
	\end{tabular}
\end{table}\noindent

The total strain energy density of a surface can be decomposed as
\eqb{lll} \label{e:E_decompose}
	\ds W_{\mathrm{total}}=W_{\mathrm{b}} + W_{\mathrm{m}}  \,, \\
\eqe
where $W_{\mathrm{b}}$ is bending energy density and $W_{\mathrm{m}}$ is membrane energy density. With the assumption of linear elastic bending behavior, the bending energy density of a sheet bent in one direction is given by
\eqb{lll} \label{e:BendingEnergyDensityShell}
\ds W_{\mathrm{b}} =  \frac{c}{2}\,\kappa^2, \\[3mm]
\eqe
where $c$ is the bending stiffness and $\kappa$ is the curvature along the bending direction. Based on (\ref{e:PolyDeformation3}) and (\ref{e:BendingEnergyDensityShell}), the total bending energy of the unit cell, therefore, follows as

\eqb{lll} \label{e:BeamEnergyCell}
\ds E_{\mathrm{beam}} = 4 b \int_{0}^{L} \frac{c}{2} \kappa^2 \dif x = \frac{24\, c\, b\,\alpha^2}{L^3}\,.\\  
\eqe

The preceding equations have been derived under the assumption of small deformations. A modified beam model suitable for larger deformations is presented in \hyperref[s:ModifiedBeamModel]{Appendix~A}.

\subsection{Structural mechanics: Rigid linkage model}\label{s:RL_Model}
\begin{figure*}[t!]
	\centering
	\begin{tabular}{c c}
		\begin{subfigure}[t]{.65\textwidth}
			\hspace*{-0.5cm}
			\centering
			\includegraphics[height=42mm]{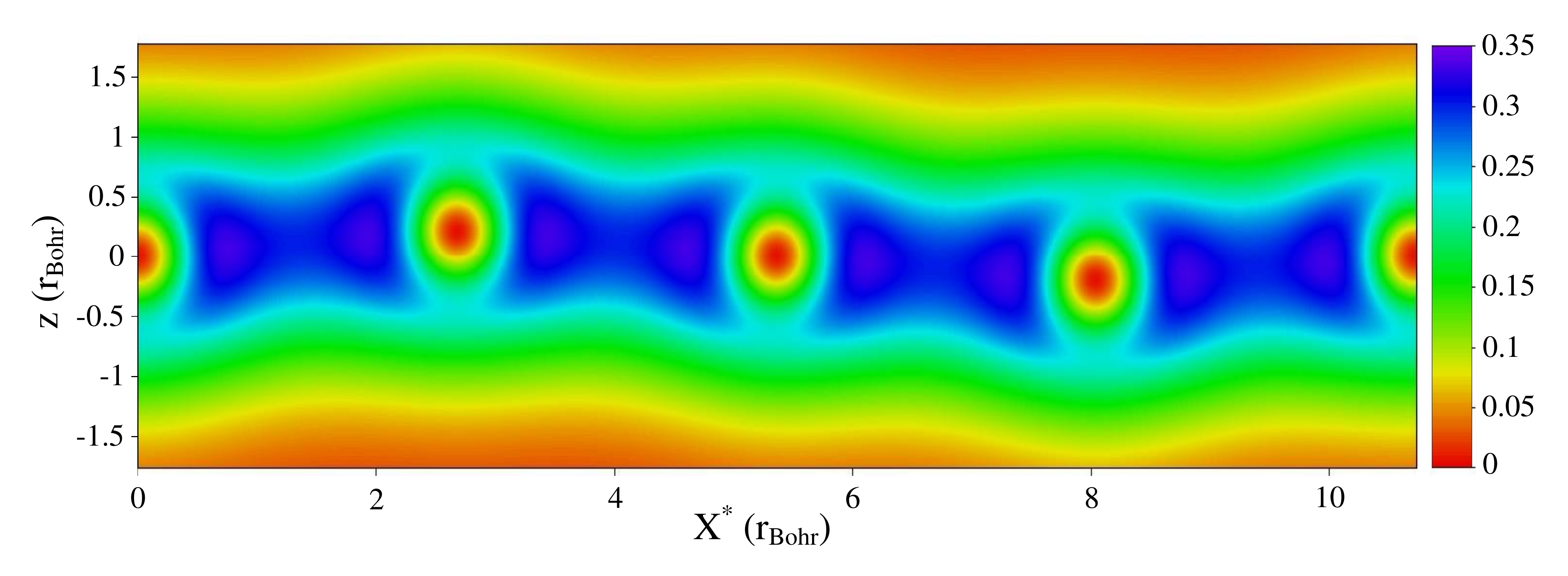}
			\caption*{(a)}
		\end{subfigure}	
		&
		\begin{subfigure}[t]{.35\textwidth}
			\centering
			\includegraphics[height=42 mm]{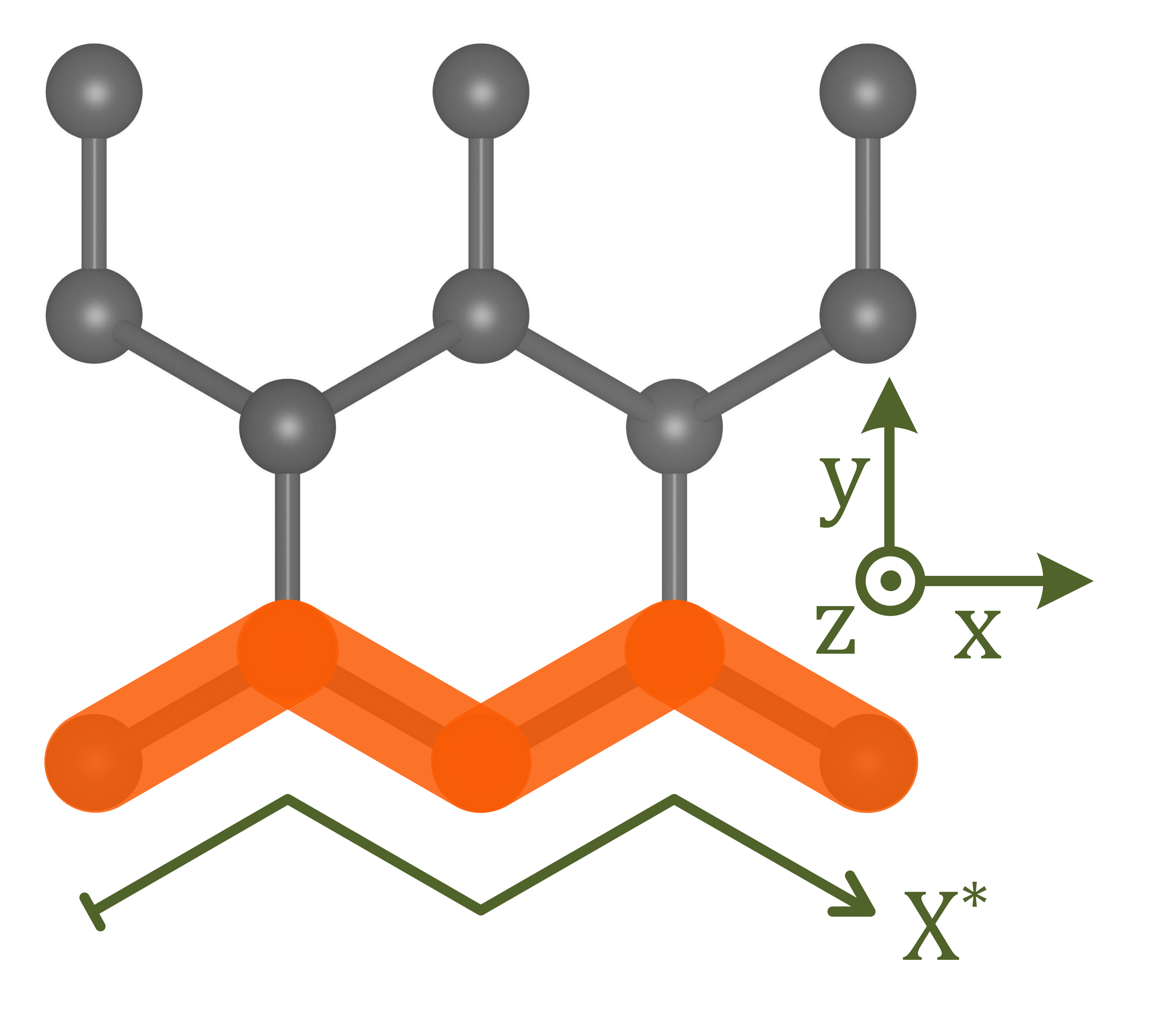}
			\caption*{(b)}
		\end{subfigure}	
	\end{tabular}
	\caption{Electron charge density ($\textrm{electron}/r_\textrm{Bohr}^3$) of the 4-atom-configuration. The charge density is plotted in (a) along the path shown in (b). Bending along the zigzag direction is shown, which is obtained by applying the out-of-plane displacement $w_0=0.207$ $r_\textrm{Bohr}$ to every second atom, as shown in Fig.~\ref{f:AllModesBC}. Plot (a) shows that the atomistic deformation of the surface resembles that of a rigid linkage.} \label{f:ElectronicDensityContourXZ} 
\end{figure*}

\begin{figure}[h] 
	\centering
	\hspace*{-0cm} 
	\includegraphics[width=90mm]{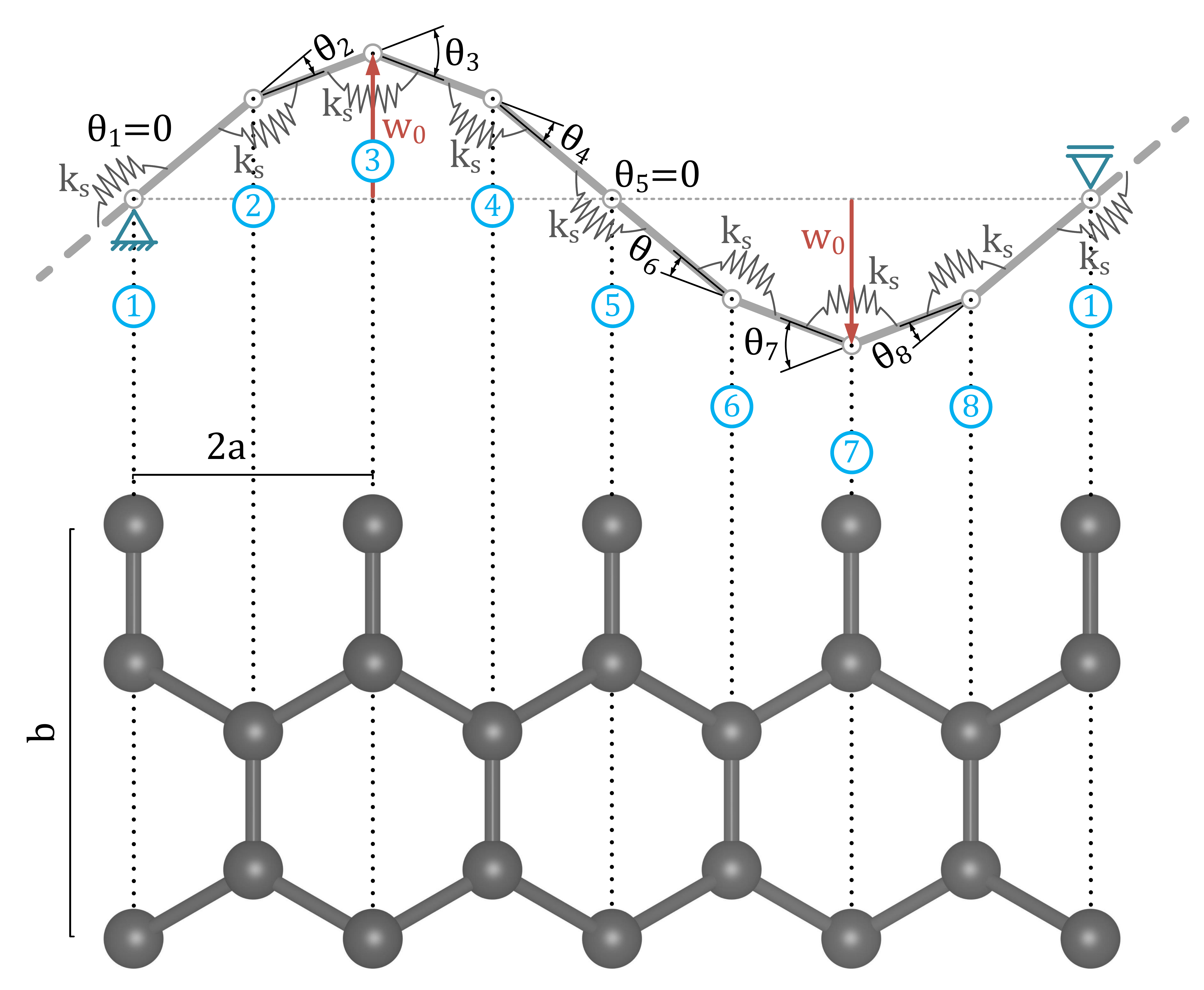}
	\vspace{0mm}
	\caption{Rigid linkage model for the 8-atom-configuration along the zigzag direction. The atomic sites are numbered in blue. For each atom $i$, $\theta_i$ denotes the out-of-plane component of the bond angle change, which is the relevant component for bending. The in-plane component is not needed, but it can be calculated from $\theta_i$, if desired.} 
	\label{f:SM}
\end{figure}

Fig.~\ref{f:ElectronicDensityContourXZ} shows the electron density of graphene along its bonds in the deformed zigzag 4-atom-configuration. The deformation resembles that of a rigid linkage -- a system of rigid rods connected by hinges and rotational springs, as shown in Fig.~\ref{f:SM}. The second method for calculating the bending stiffness thus uses rigid linkage theory. In such a system, the rods (bonds) do not contribute to bending, and the bending energy is exclusively stored in the rotational springs (around the atoms), where the angle changes between the rods (bonds) occur. This assumption is valid for most cases. However, in some rare cases, the electron density of the C-C bond does not resemble a straight line \citep{Wiberg1996_01,Doedens2017_01}.

In the linear regime, the energy stored in a rotational spring is given by 
\eqb{lll} \label{e:RotationalSpringEnergy0}
	\ds U_{\mathrm{s}} =  \frac{k_{\mathrm{s}}\, b}{2}\,{\theta}^2\,, \\[3mm]
\eqe
where the constant $k_{\mathrm{s}}$ is the rotational spring stiffness per width of the unit cell $b$, and $\theta$ is the rotation angle calculated from the current relaxed position of the adjacent atoms. The total rotational spring energy of a unit cell consisting of $n$ atoms then becomes
\eqb{lll}\label{e:Ecellspring}
\ds E_{\mathrm{linkage}} = \frac{k_{\mathrm{s}}\,b}{2} \sum_{i=1}^{n}  \theta_{i}^2 \,, \\[3mm]
\eqe
where $\theta_i$ denote all the angles between the linkages as shown in Fig.~\ref{f:SM}. These angles can be taken directly from DFT. Alternatively, they can be determined analytically for a given $w_0$ by calculating the atomic displacements $w_i$ of all the free atoms (atoms 2, 4, 6, and 8 in Fig.~\ref{f:SM}) from the condition that their force $F_i:= \partial E_{\mathrm{linkage}} / \partial w_i$ vanishes. Substituting the obtained result for all $\theta_i$ back into Eq.~(\ref{e:Ecellspring}), while assuming small deformations ($\tan{\theta} \approx \theta$), then gives 
\eqb{lll}\label{e:EcellspringSimplified}
\ds E_{\mathrm{linkage}} = \frac{k_{\mathrm{s}}\,b\, \xi}{L^2} w_0^2 \,, \\[3mm]
\eqe
where $\xi$ is a constant for each atomic configuration listed in Tab.~\ref{t:alfas}.

As discussed below, $k_{\mathrm{s}}$ is calibrated from atomistic data. Once this is done, Eq.~(\ref{e:Ecellspring}) or (\ref{e:EcellspringSimplified}) can be used to calculate the energy of the deformed linkage. Since the spring constant $k_\mathrm{s}$ is not the same as the bending stiffness $c$ of the continuum model, the relation between $k_\mathrm{s}$ and $c$ is needed in order to determine $c$ from the linkage model. As shown in \hyperref[s:assessment_c]{Appendix~B}, the equivalent bending stiffness associated with the rotational spring stiffness of the rigid linkage model is    
\eqb{lll} \label{e:cK_relation}
\ds c =  k_{\mathrm{s}} \bar{a}\,, \\ [3mm]
\eqe
where $\bar a$ is the average spacing between adjacent atoms that can be calculated by dividing $4L$ by the atomic configuration number. With this, the energies of the unit cells given in Eqs.~(\ref{e:Ecellspring}) and (\ref{e:EcellspringSimplified}) become
\eqb{lll} \label{e:SpringEnergyCell}
\ds E_{\mathrm{linkage}} = \frac{b\, c}{2 \bar{a}} \sum_{i=1}^{n}  \theta_{i}^2 \\[3mm]
\eqe
in general, and 
\eqb{lll}\label{e:SpringEnergyCellSimplified}
\ds E_{\mathrm{linkage}} = \frac{b\, c\, \xi}{\bar{a}\, L^2} w_0^2  \\[3mm]
\eqe
for infinitesimal strains.

\subsection{Calculation of the bending stiffness}
In a relaxed unit cell with the described loading conditions, the membrane energies are negligible at small deformations, which is confirmed by the following results. Therefore, according to Eq.~(\ref{e:E_decompose}), the total energy from the DFT calculations corresponds to bending energy.
The bending stiffness $c$ can then be calculated from the DFT data by minimizing the cost function
\eqb{lll}
\ds C_2(c) = \sum_{k=1}^{N_{p} }  \bigg(E^{(k)}_{\mathrm{model}}(c)-E^{(k)}_{\mathrm{DFT}}\bigg)^2 \,, 
\label{e:Cost2}
\eqe
where $k$ denotes the load increment, $N_{p} $ is the total number of load increments for each atomic configuration, $E^{(k)}_{\mathrm{DFT}}$ are the energies from DFT calculations, and $E^{(k)}_{\mathrm{model}}$ are the energies of the unit cell predicted by either Eq.~(\ref{e:BeamEnergyCell}) for the beam model, or Eqs.~(\ref{e:SpringEnergyCell}) and (\ref{e:SpringEnergyCellSimplified}) for the rigid linkage model. The minimization of the cost function~(\ref{e:Cost2}) results in one value for the bending stiffness for each atomic configuration.

\section{Results} \label{s:Results}
This section presents the results for the bending stiffness from different atomic configurations. First, graphene in the small deformation regime is examined (Sec.~\ref{s:R_Validation_Graphene}). This is followed by graphene at moderately large deformations (Sec.~\ref{s:R_Large_Graphene}). Finally, h-BN, silicene, and blue phosphorene in the small deformation regime are examined (Sec.~\ref{s:R_other_materials}). The new results are compared with existing results from the literature.\\ 
Small deformations are ensured by picking an average curvature that is at least 100 times smaller than the inverse of the lattice parameter ($\bar \kappa < 0.01 / a$). For the considered moderately large deformations, on the other hand, the average curvature can be as large as 2 times the inverse lattice parameter ($\bar \kappa < 0.5 / a$). In all cases the loading was found to be reversible, implying that the material behavior is elastic in the studied ranges.

\subsection{Calibration and validation for graphene at small bending deformations} \label{s:R_Validation_Graphene}
\begin{figure}
	\centering
	\begin{subfigure}[b]{0.45\textwidth}
		\centering
		\includegraphics[width=\textwidth]{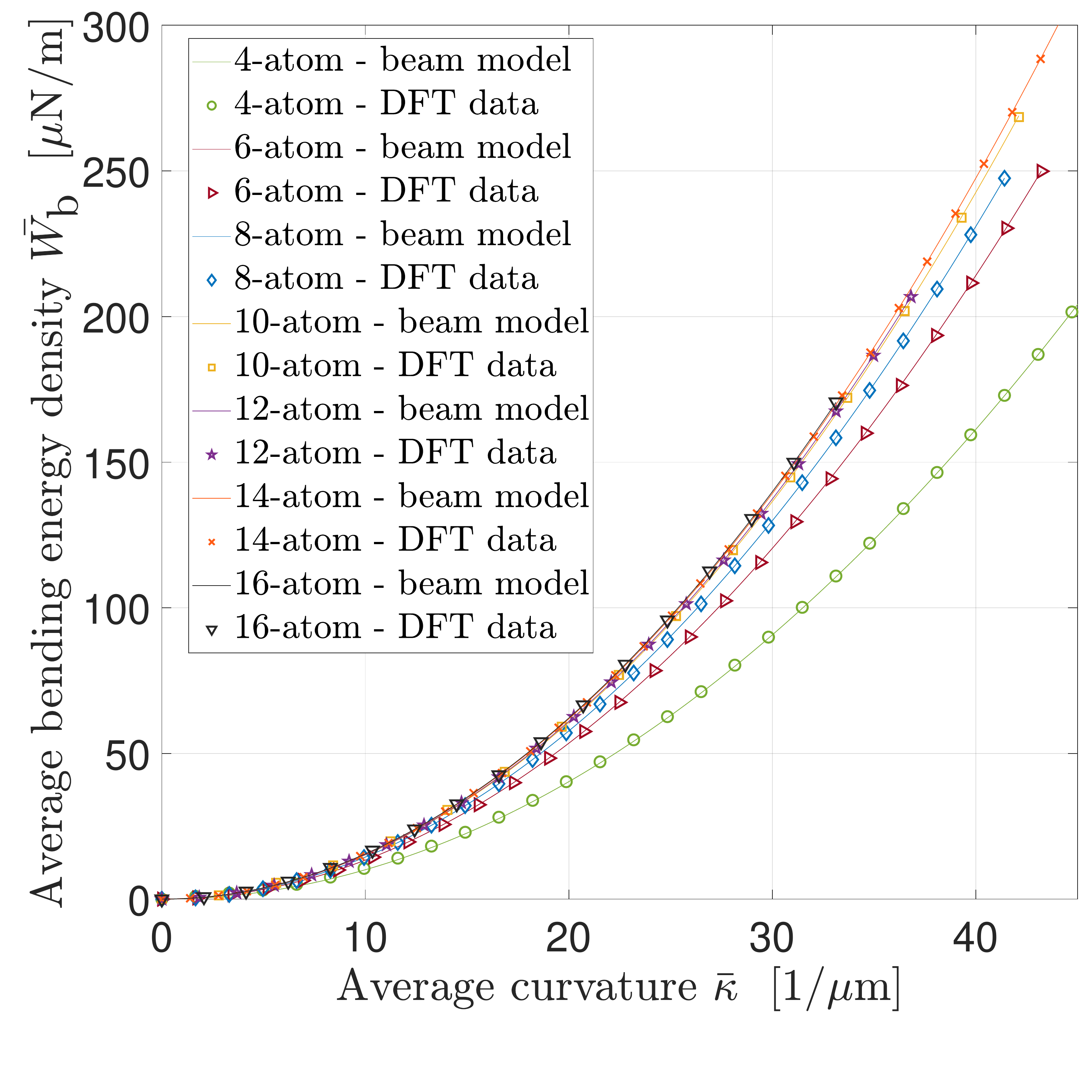}
		\vspace{-11mm}
		\caption{Bending calibration in the zigzag direction using the beam model}
		\label{f:EnergiesVSkappaPolyZigzag}
	\end{subfigure}
	\hspace*{0.05\textwidth}
	\begin{subfigure}[b]{0.45\textwidth}
		\centering
		\includegraphics[width=\textwidth]{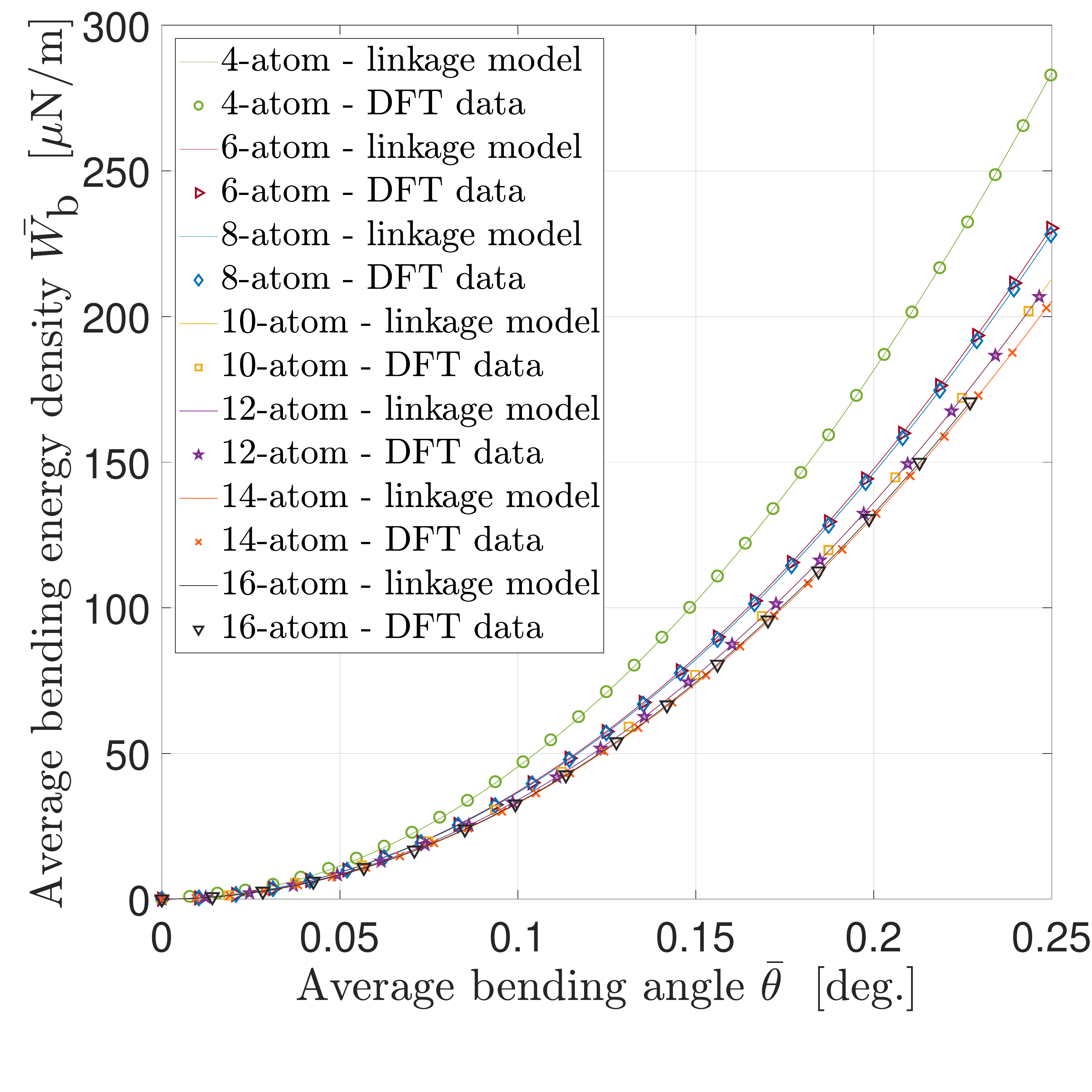}
		\vspace{-11mm}
		\caption{Bending calibration in the zigzag direction using the rigid linkage model}
		\label{f:EnergiesVSkappaSpringZigzag}
	\end{subfigure}
	
	\begin{subfigure}[b]{0.45\textwidth}
		\centering
		\vspace{5mm}
		\includegraphics[width=\textwidth]{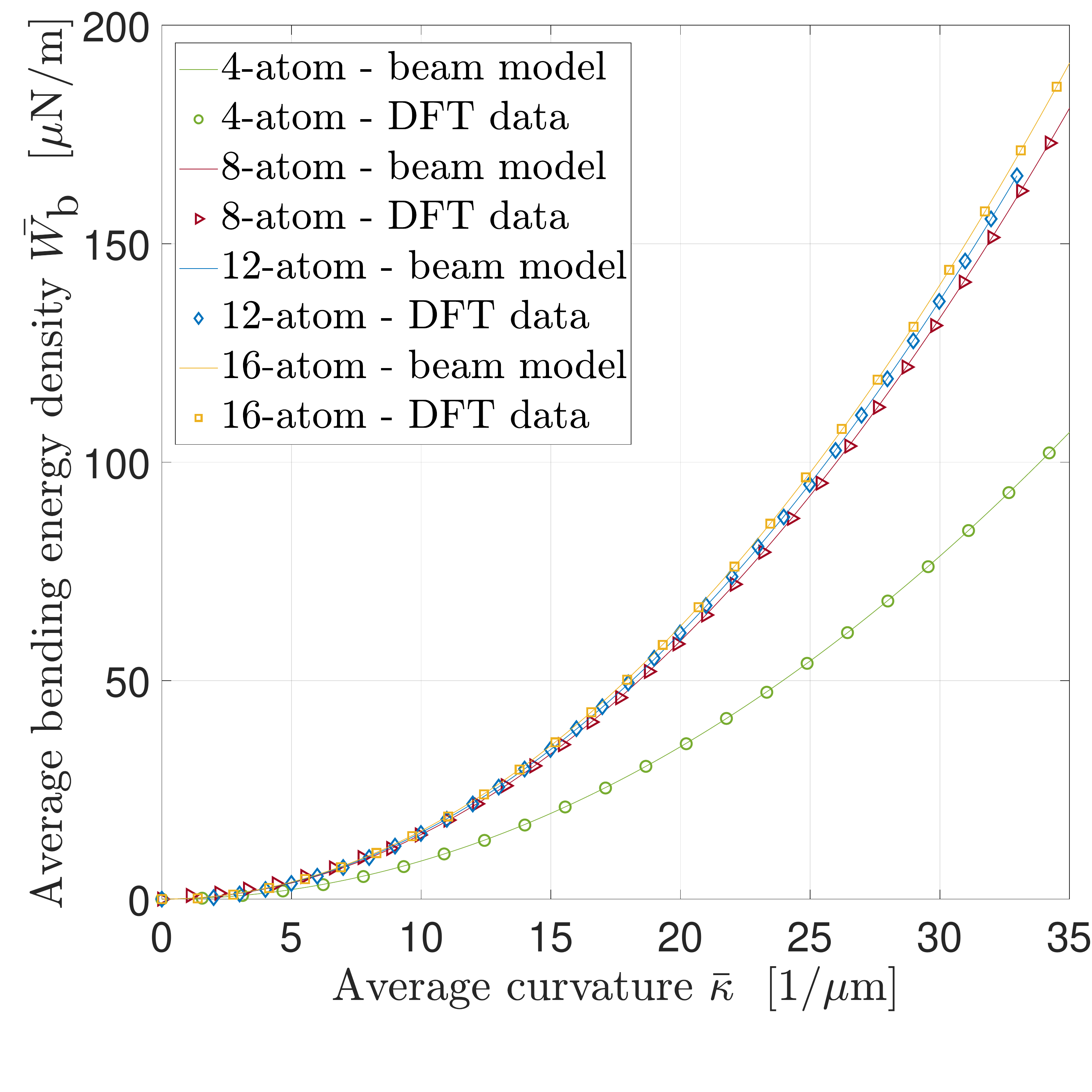}
		\vspace{-11mm}
		\caption{Bending calibration in the armchair direction using the beam model}
		\label{f:EnergiesVSkappaPolyArmchair}
	\end{subfigure}
	\hspace*{0.05\textwidth}
	\begin{subfigure}[b]{0.45\textwidth}
		\centering
		\vspace{5mm}
		\includegraphics[width=\textwidth]{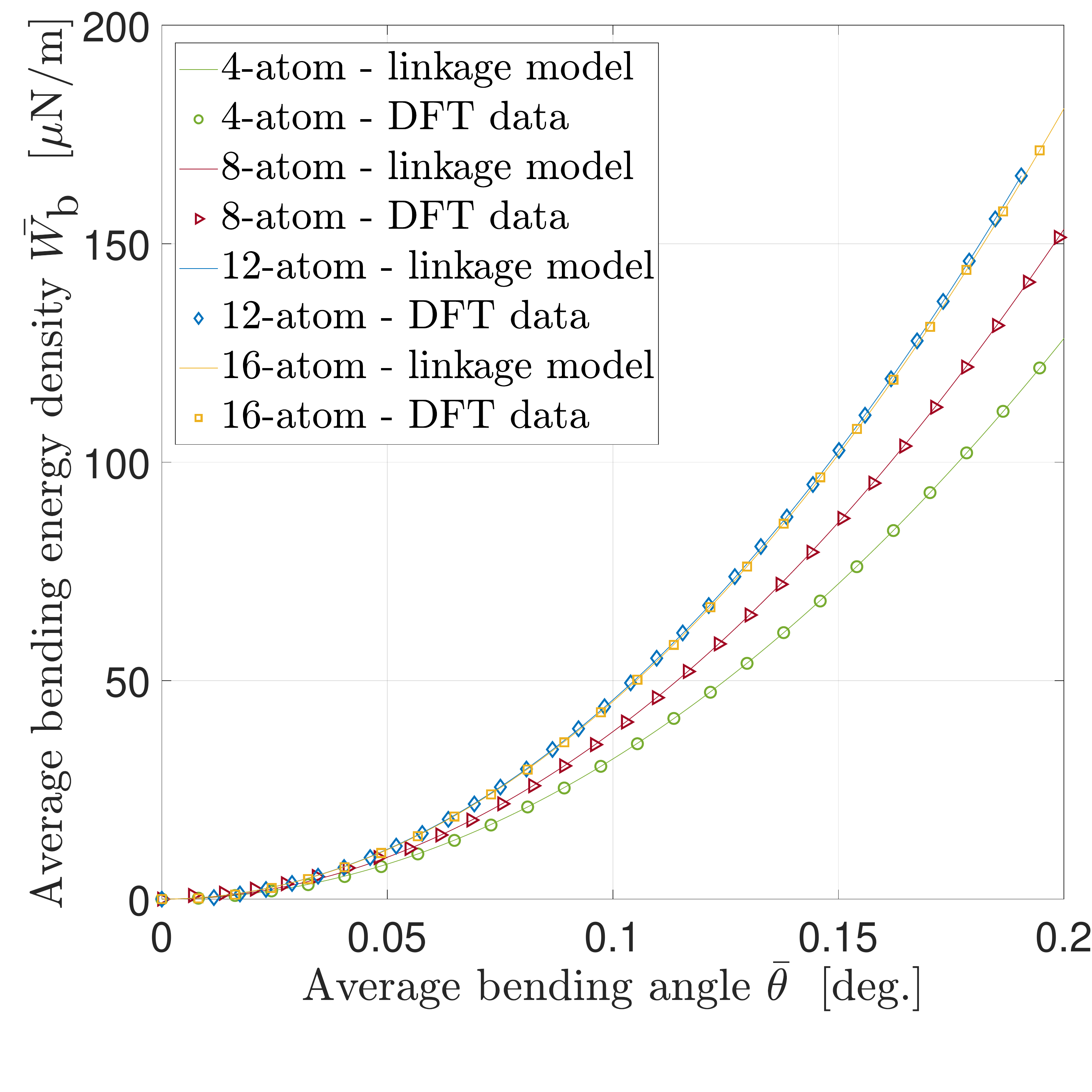}
		\vspace{-11mm}
		\caption{Bending calibration in the armchair direction using the rigid linkage model}
		\label{f:EnergiesVSkappaSpringArmchair}
	\end{subfigure}
	\caption{Atomistic data and calibrated bending models for graphene at small deformations. As seen, the proposed models describe the atomistic bending behavior very well. The plots show the average bending energy density versus the average curvature $\bar{\kappa}$ for the beam model in (a) and (c), and versus the average bending angle $\bar{\theta}$ for the rigid linkage model in (b) and (d). The DFT data in (a) and (b) as well as in (c) and (d) are the same, respectively. It is just plotted versus the different kinematic quantities ($\bar\kappa$ and $\bar\theta$) of the two models.}
	\label{f:EnergiesVSkappaAllGraphene}
\end{figure}

Fig.~\ref{f:EnergiesVSkappaAllGraphene} shows the DFT results for the average bending energy density of the unit cell -- i.e., the bending energy divided by the surface area -- together with the corresponding results for the calibrated beam and linkage models. The results are presented as a function of the average curvature $\bar{\kappa}$ and average angle change $\bar{\theta}$, respectively, for both the zigzag and the armchair directions. The averages $\bar\kappa$ and $\bar\theta$ are defined in \hyperref[s:Supplementary]{Appendix~C}. 
The calculated bending stiffnesses from the calibrated models are shown in Tab.~\ref{t:cMeanStdv}, together with reference results. In order to calculate the bending stiffnesses for the linkage model, first, the spring constants are calculated and then translated to bending stiffnesses using Eq.~(\ref{e:cK_relation}). 
\begin{table}[]
	\centering
	\small
	\caption{Bending stiffness [nN.nm] of graphene according to the beam and linkage model using the different atomic configurations from Fig.~\ref{f:AllModesBC}. The values can be expected more accurate the larger the unit cell, and more accurate for the linkage model. Existing values from the literature are also shown.} \label{t:cMeanStdv}
	\vspace{2mm}
	\begin{tabular}{lcccc} 
		\hline \\ [-4mm]
		 &\multicolumn{2}{c}{Zigzag direction} & \multicolumn{2}{c}{Armchair direction}\\  
		\hline \\ [-4mm]
		 Unit cell & Beam model & Linkage model & Beam model & Linkage model \\ 
		\hline \\ [-4mm]
		4-atom-config. & 0.1512 & 0.2268 & 0.1308  & 0.1923  \\
		\hline \\ [-4mm]
		6-atom-config. & 0.2007 & 0.2459 &   &          \\
		\hline \\ [-4mm]
		8-atom-config. & 0.2165 & 0.2436 & 0.2217 & 0.2257  \\
		\hline \\ [-4mm]
		10-atom-config. & 0.2244 & 0.2418 &   & \\
		\hline \\ [-4mm]
		12-atom-config. & 0.2289  & 0.2417 & 0.2284 & 0.2468 \\
		\hline \\ [-4mm]
		14-atom-config. & 0.2321 & 0.2411 &   &        \\
		\hline \\ [-4mm]
		16-atom-config. & 0.2331  & 0.2405 & 0.2343 &  0.2441 \\
		\hline
		\hline\\ [-4mm]
		\multicolumn{2}{l}{Other works} & Zigzag & No distinction & Armchair \\
		\hline \\ [-4mm]
		\multicolumn{2}{l}{\citet{Kudin2001_01}} & & 0.2385 & \\
		\hline \\ [-4mm]
		\multicolumn{2}{l}{\citet{Lu2009_01}} & & 0.225 & \\
		\hline \\ [-4mm]
        \multicolumn{2}{l}{\citet{Wei2013_01}} & & 0.231 & \\
		\hline \\ [-4mm]
		\multicolumn{2}{l}{\citet{Kumar2020_01}} & 0.2419 & & 0.2403\\
		\hline
	\end{tabular}
\end{table}\noindent

\subsection{Graphene at moderately large bending deformations} \label{s:R_Large_Graphene}
\begin{figure}[h]
	\centering
	\begin{subfigure}[b]{0.45\textwidth}
		\centering
		\includegraphics[width=\textwidth]{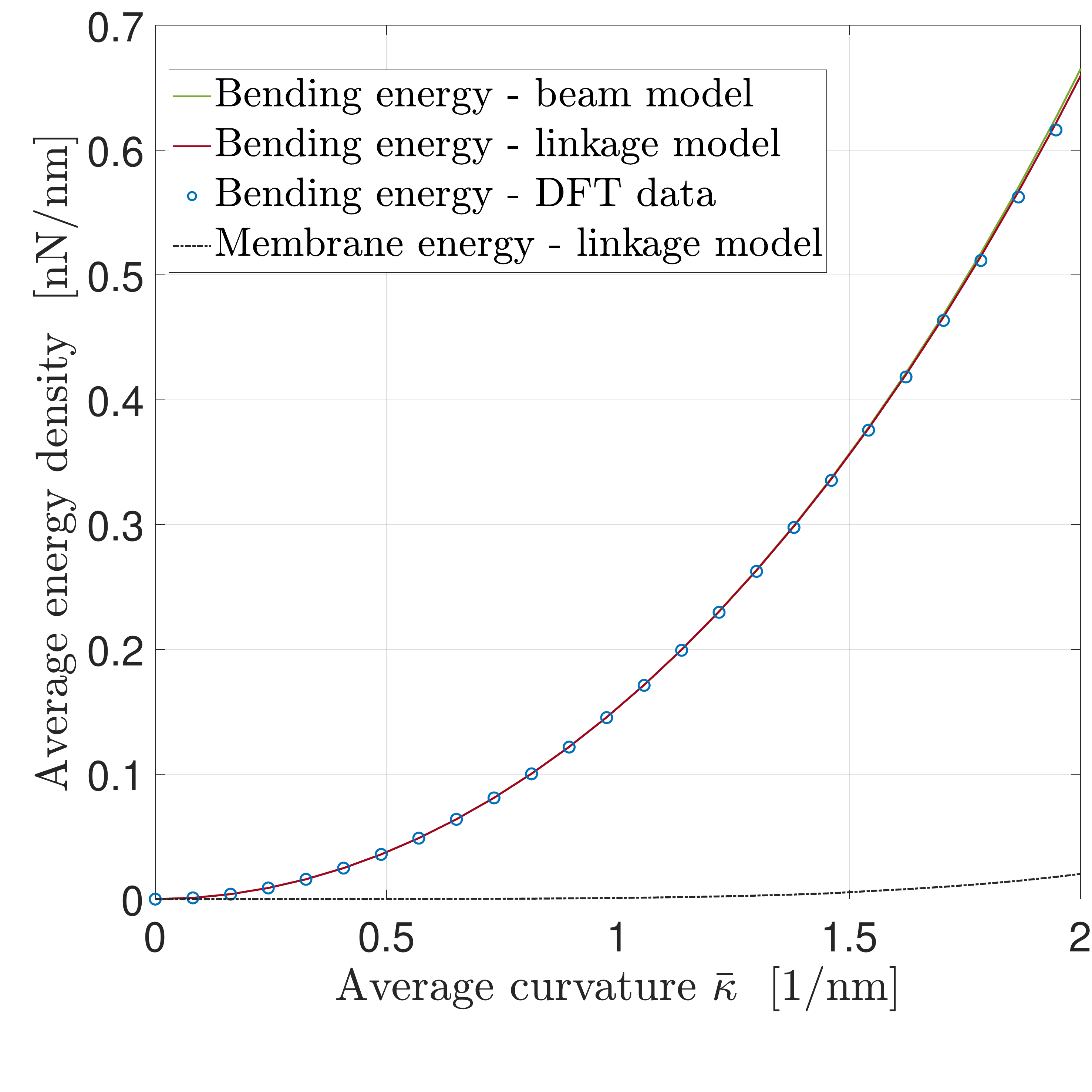}
		\vspace{-9mm}
		\caption{8-atom zigzag configuration}
		\label{f:8Z-Large}
	\end{subfigure}
	\hspace*{ 0.05\textwidth}
	\begin{subfigure}[b]{0.45\textwidth}
		\centering
		\includegraphics[width=\textwidth]{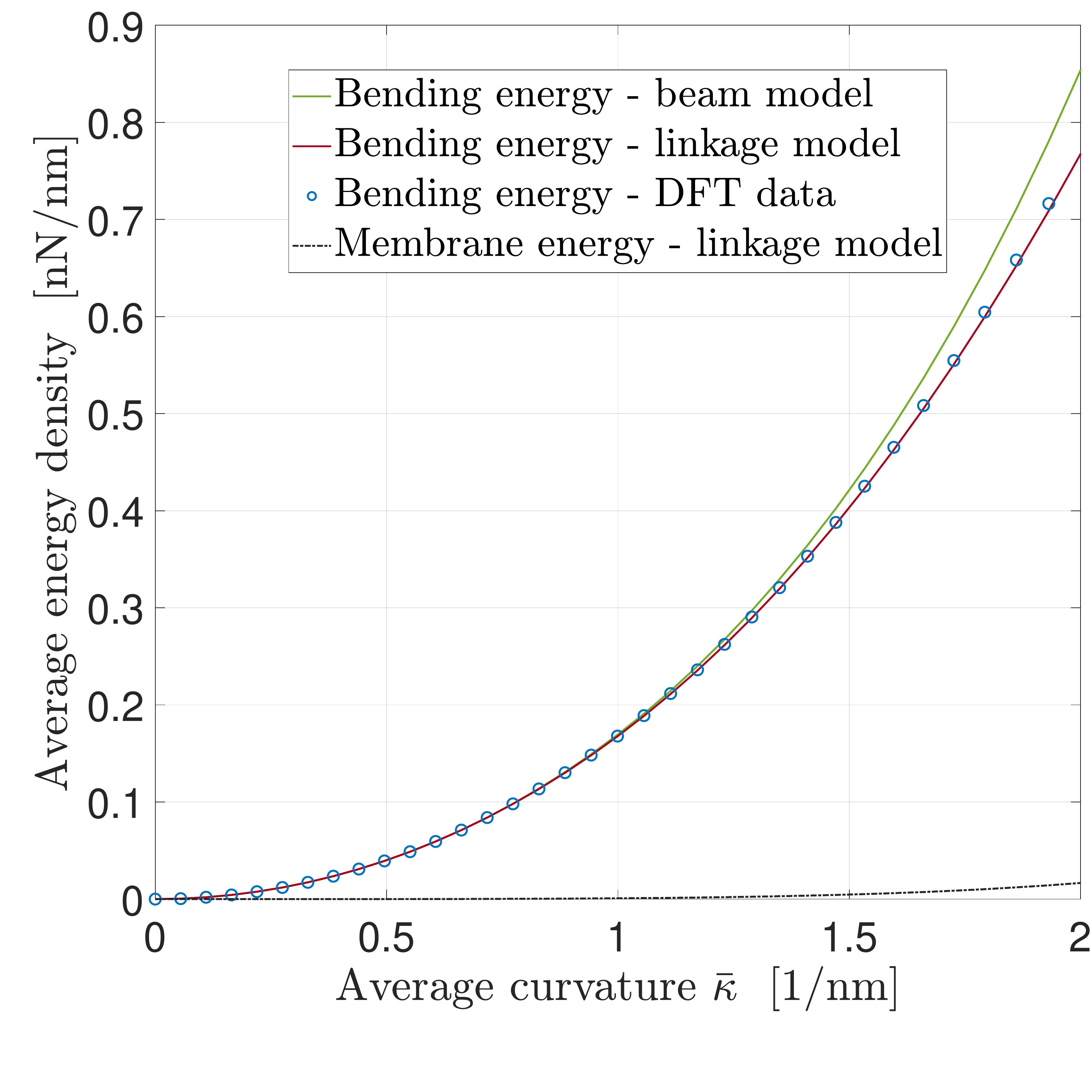}
		\vspace{-9mm}
		\caption{16-atom armchair configuration}
		\label{f:16A-Large}
	\end{subfigure}
	\vspace{0 mm}
	\caption{Beam and linkage models for graphene at moderately large deformations. Shown are the 8-atom zigzag and 16-atom armchair configuration. As seen, the proposed calibration still works very well. This is because the membrane energies are still negligible compared to the bending energies, demonstrating that the problem is still bending dominated at these moderately large curvatures.}  \label{f:EnergiesVSkappaLarge}
\end{figure}

For the calibration of the bending stiffness based on the models presented above, it is necessary to apply infinitesimal strains. At large deformations, however, the bending stiffness is generally no longer constant, such that these models can become inaccurate. In order to investigate how this affects graphene, we assess the performance of the calibrated bending models of Tab.~\ref{t:cMeanStdv}
for moderately large deformations. The 8-atom zigzag and 16-atom armchair configurations are considered here, and the applied curvatures are almost two orders of magnitude larger than for the small deformation cases of Tab.~\ref{t:cMeanStdv}. The results are presented in Fig.~\ref{f:EnergiesVSkappaLarge}. In both tests, it was observed that graphene expands along the $y$-direction as a result of bending.\\
In order to accurately calculate the bending energies for larger deformations, some of the simplifications for infinitesimal deformations need to be reconsidered. Eqs.~(\ref{e:BeamEnergyCell}) and (\ref{e:SpringEnergyCellSimplified}) have been developed for infinitesimal strains for the beam and linkage models, respectively. Therefore, they are not suitable for the calculation of the bending energy when atomistic deformations are moderately large. For the beam model, the energy can be evaluated more accurately if the exact curvature expression is used, as discussed in \hyperref[s:ModifiedBeamModel]{Appendix~A}. For the rigid linkage model, the more general expressions (\ref{e:Ecellspring}) and (\ref{e:SpringEnergyCell}), which are not restricted to small deformations, have to be used instead of (\ref{e:EcellspringSimplified}) and (\ref{e:SpringEnergyCellSimplified}). Additionally, Eq.~(\ref{e:SpringEnergyCell}) should now be evaluated by accounting for the length changes in $L$ and $b$, obtained from the DFT calculations.\\
Another aspect that needs to be considered at large deformations is the contribution of the membrane strain energy to the total energy, even though the atomistic tests are designed to minimize this energy. The calculation of the membrane energy at each load increment based on the linkage model is discussed in \hyperref[s:MembraneEnergy]{Appendix~D}. The calculated membrane energies are also shown in Fig.~\ref{f:EnergiesVSkappaLarge}.
\subsection{Application to other 2D materials} \label{s:R_other_materials}
\begin{figure}
	\centering
	\begin{subfigure}[t]{0.45\textwidth}
		\centering
		\includegraphics[width=\textwidth]{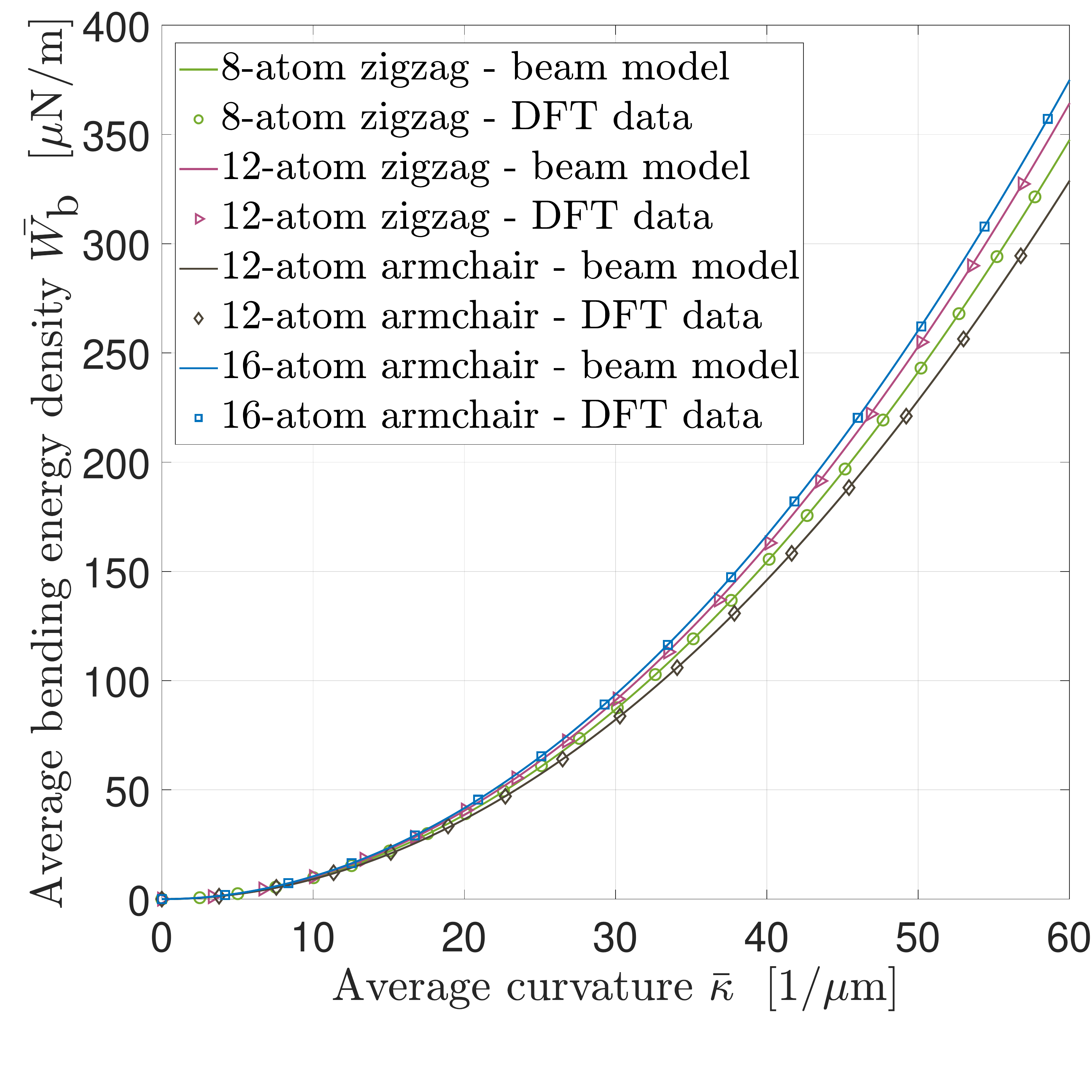}
		\vspace{-11mm}
		\caption{Calibration of h-BN from the beam model}
		
		\label{f:BNBeam}
	\end{subfigure}
	\hspace*{0.05\textwidth}
	\begin{subfigure}[t]{0.45\textwidth}
		\centering
		\includegraphics[width=\textwidth]{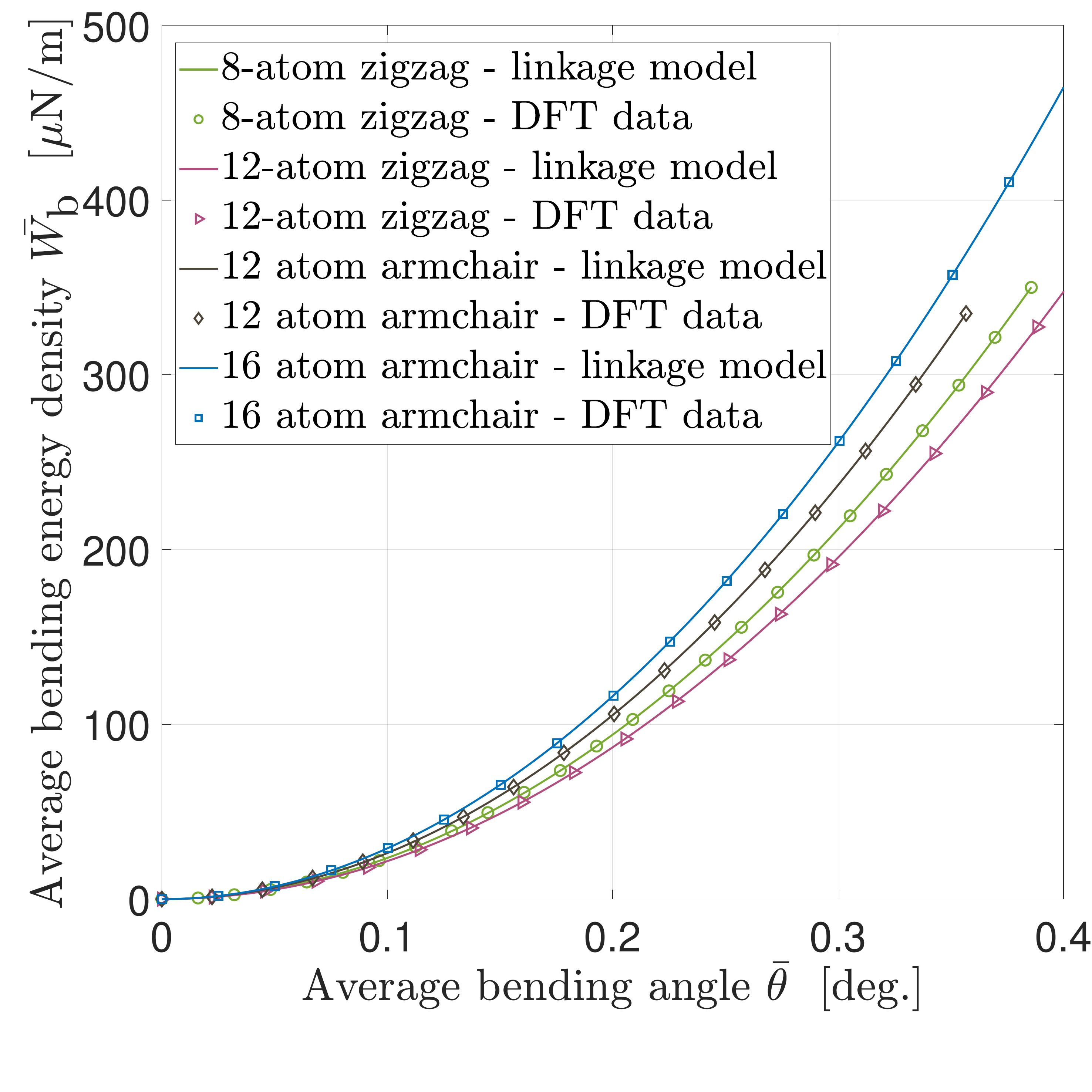}
		\vspace{-11mm}
		\caption{Calibration of h-BN from the linkage model}
		\label{f:BNSpring}
	\end{subfigure}
	\\
	\begin{subfigure}[t]{0.45\textwidth}
		\centering
		\includegraphics[width=\textwidth]{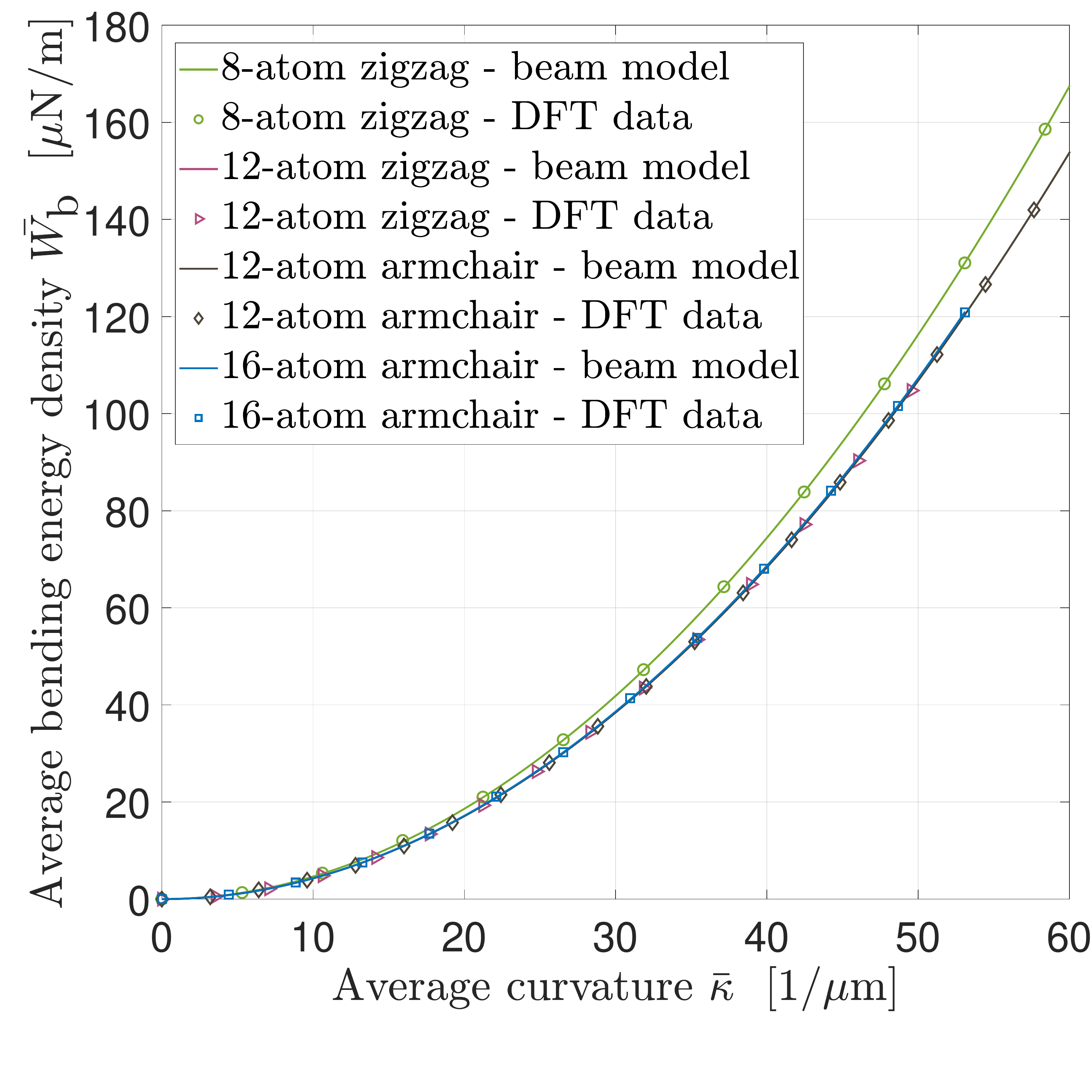}
		\vspace{-11mm}
		\caption{Calibration of silicene from the beam model}
		\label{f:SiBeam}
	\end{subfigure}
	\hspace*{0.05\textwidth}
	\begin{subfigure}[t]{0.45\textwidth}
		\centering
		\includegraphics[width=\textwidth]{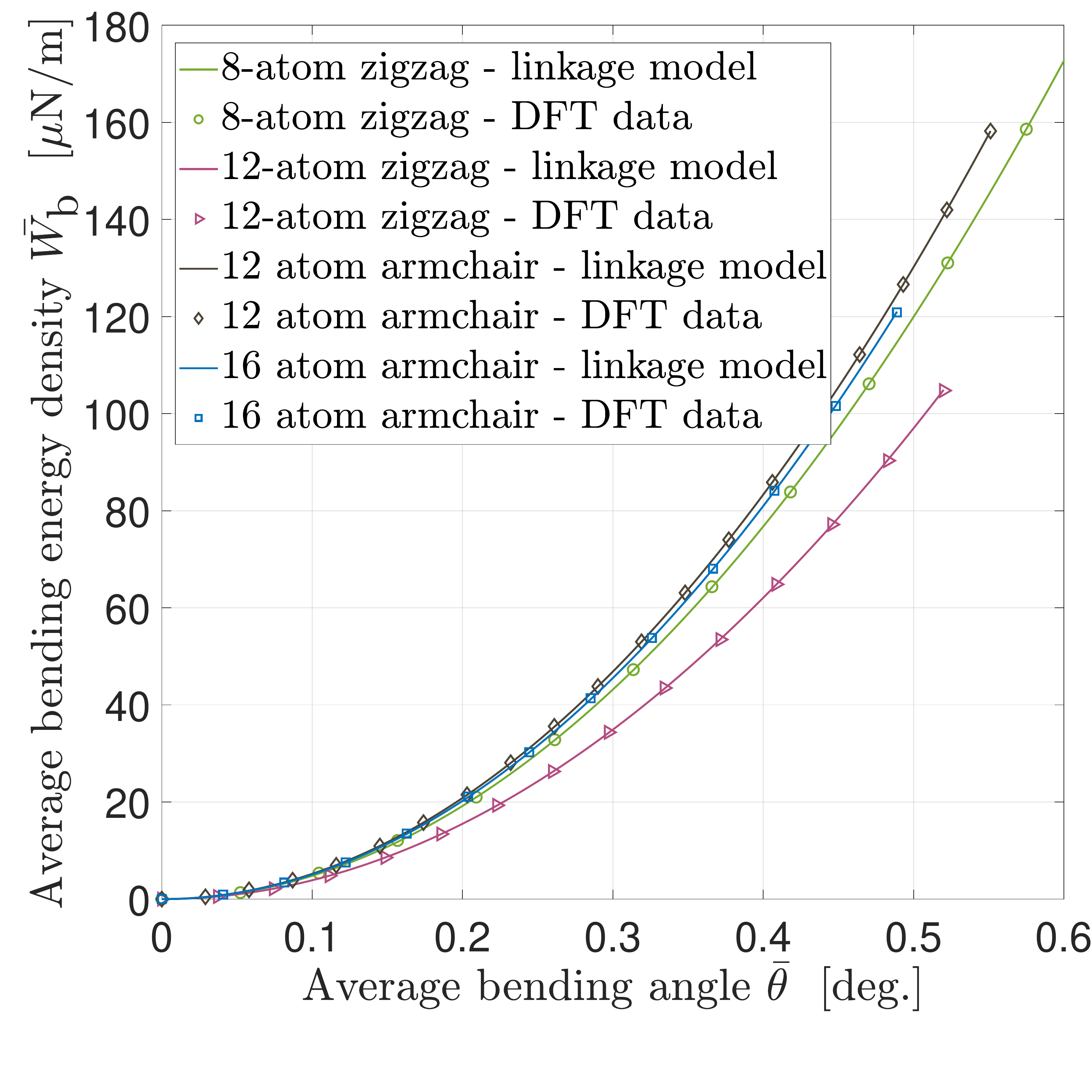}
		\vspace{-11mm}
		\caption{Calibration of silicene from the linkage model}
		\label{f:SiSpring}
	\end{subfigure}
	\\
	\begin{subfigure}[t]{0.45\textwidth}
		\centering
		\includegraphics[width=\textwidth]{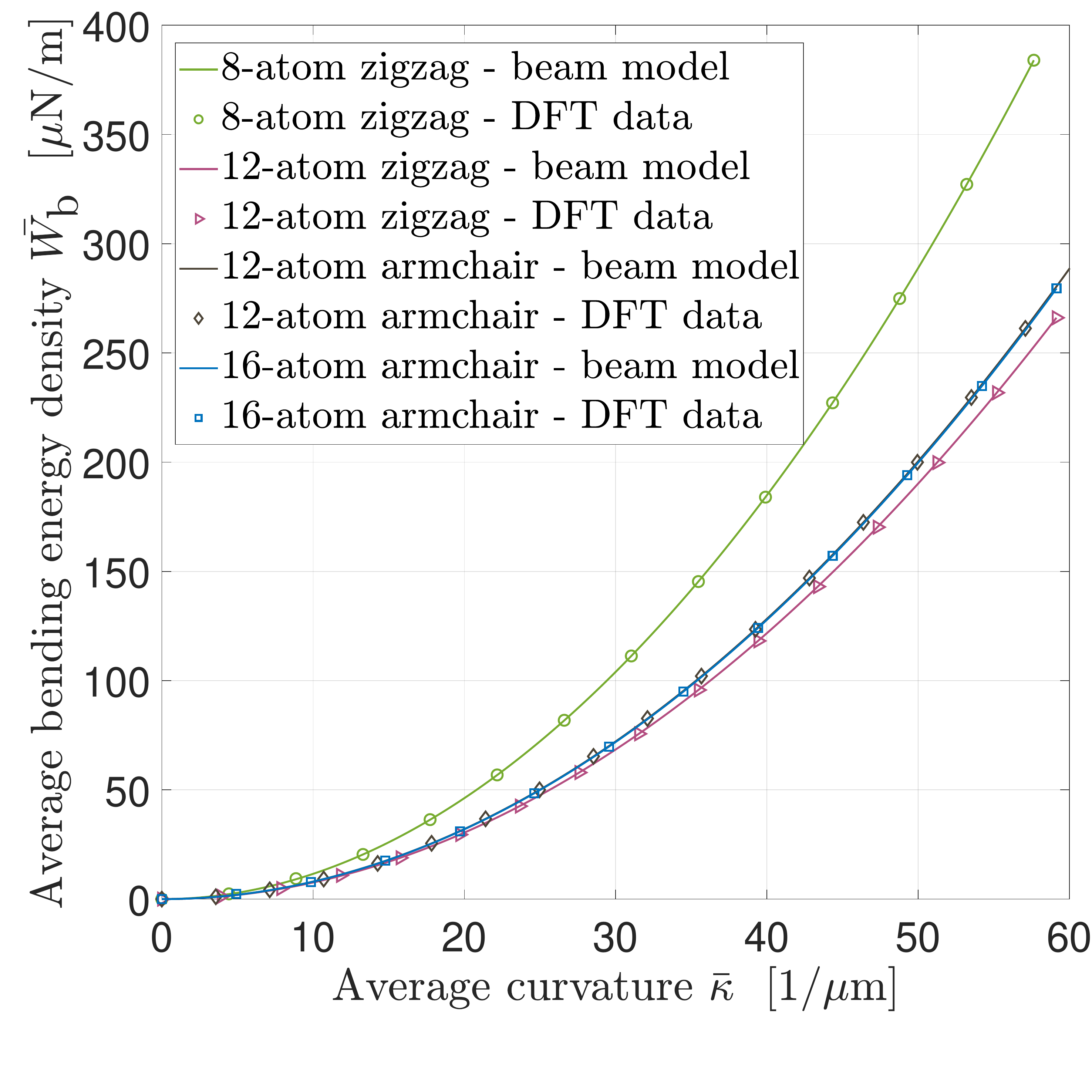}
		\vspace{-11mm}
		\caption{Calibration of blue P from the beam model}
		\label{f:BluePBeam}
	\end{subfigure}
	\hspace*{0.05\textwidth}
	\begin{subfigure}[t]{0.45\textwidth}
		\centering
		\includegraphics[width=\textwidth]{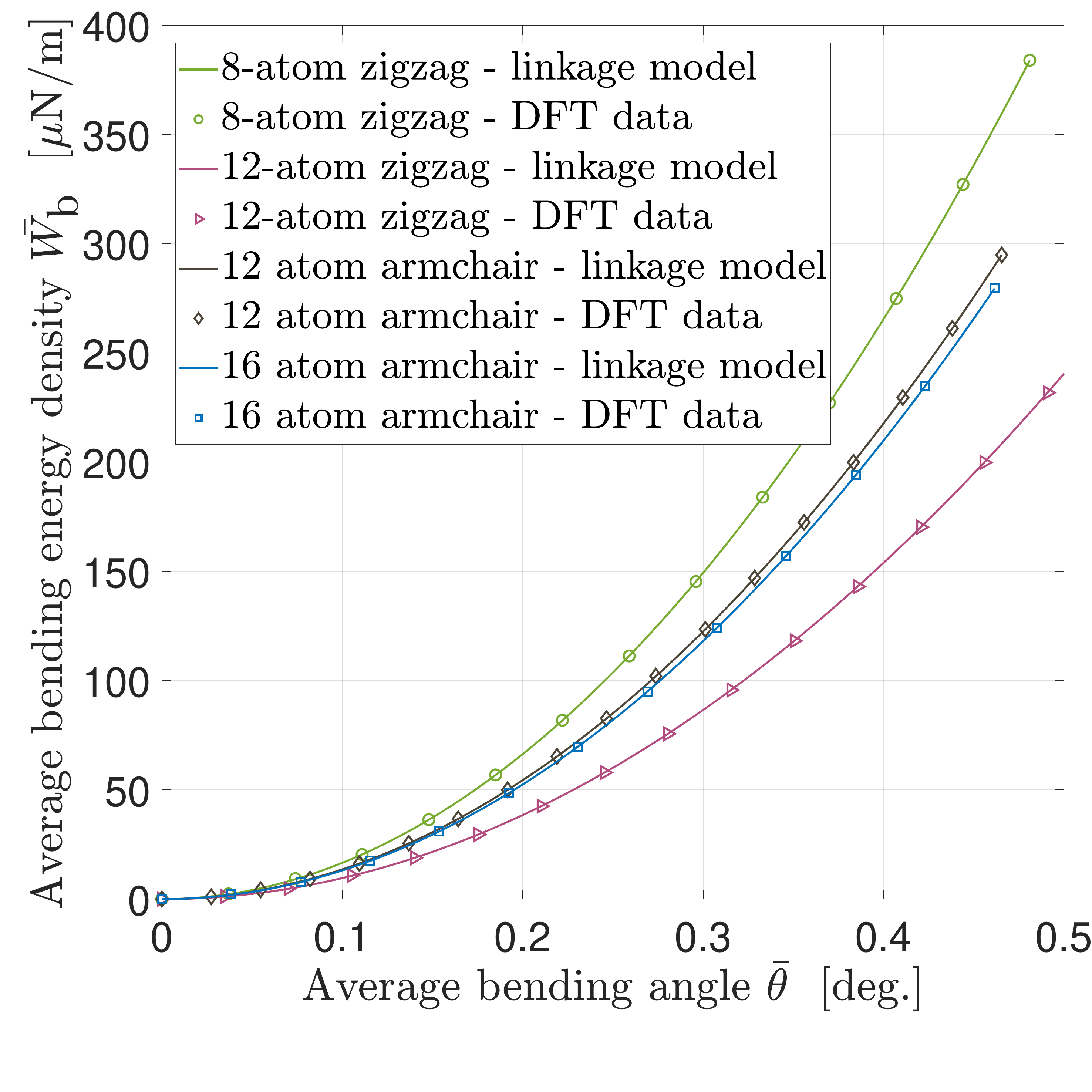}
		\vspace{-11mm}
		\caption{Calibration of blue P from the linkage model}
		\label{f:BluePSpring}
	\end{subfigure}
	\vspace{-3mm}
	\caption{Atomistic data and calibrated models for hexagonal boron nitride, silicene, and blue phosphorene at small deformations. Left side: beam model calibration. Right side: linkage model calibration. Again, the proposed models describe the atomistic bending behavior very well. The DFT data in the corresponding left and right figures are the same for each material. It is just plotted versus the different kinematic quantities ($\bar\kappa$ and $\bar\theta$) of the two models.
	}
	\label{f:BNSiliceneEPlot}
\end{figure}

The proposed approach can also be used to calculate the bending stiffness for other 2D materials. This is demonstrated on hexagonal boron-nitride (h-BN), silicene, and blue phosphorene.  Fig.~\ref{f:BNSiliceneEPlot} shows their bending calibration along the zigzag and armchair directions. For the zigzag direction, the 8- and 12-atom-configurations, and for the armchair direction, the 12- and 16-atom-configurations are considered.\\
The determined values for the bending stiffness of h-BN, silicene, and blue phosphorene are listed in Tab.~\ref{t:cAll}, along with the already determined value of graphene at small deformations and known values from the literature. The reported bending stiffnesses for graphene correspond to the 16-atom-configurations along the zigzag and armchair directions from Tab.~\ref{t:cMeanStdv}. For h-BN, silicene, and blue phosphorene, the reported values correspond to the 12-atom-configurations along the zigzag direction and the 16-atom-configurations along the armchair direction. The values of the bending stiffness are reported separately for each atomic configuration in Tab.~\ref{t:BendingStiffnessOthers}.
One additional parameter of corrugated lattices such as silicene and blue phosphorene in comparison to flat ones such as graphene and h-BN is their initial thickness. This is accounted for in our approach by applying the displacement $w_0$ directly to the initial position of the atoms. This is equivalent to assuming $w_0$ is applied to the midsurface of the corrugated material and further assuming that lateral displacements of non-midsurface atoms as well as thickness changes are negligible during deformation, which is reasonable for small deformations.
\defcitealias{Wei2013_01}{W13}
\defcitealias{Kudin2001_01}{K01}
\defcitealias{Lu2009_01}{L09}
\defcitealias{Kumar2020_01}{K20}
\defcitealias{Qu2019_01}{Q19}
\defcitealias{Roman2014_01}{R14}
\defcitealias{Banerjee2016_01}{B16}
\defcitealias{Zhang2020_01}{Z20}
\begin{table}[H]
	\centering
	\caption{Calibrated bending stiffnesses [nN.nm] of the investigated 2D materials and their comparison with reference values from the literature, namely
    \citetalias{Kudin2001_01} \citep{Kudin2001_01},
    \citetalias{Lu2009_01} \citep{Lu2009_01},
    \citetalias{Wei2013_01} \citep{Wei2013_01},
    \citetalias{Roman2014_01} \citep{Roman2014_01},
    \citetalias{Banerjee2016_01} \citep{Banerjee2016_01},
    \citetalias{Qu2019_01} \citep{Qu2019_01},
    \citetalias{Kumar2020_01} \citep{Kumar2020_01}, and
    \citetalias{Zhang2020_01} \citep{Zhang2020_01}. The calibrated bending stiffnesses are from the largest available unit cell in all cases (see Fig.~\ref{f:BNSiliceneEPlot}), as it can be expected to give the most accurate values. In the cited literature with only one value, no distinction between zigzag and armchair directions is made. All reference results are based on ab-initio calculations, except the large silicene value of \citet{Roman2014_01}, which is based on MD calculations.} \label{t:cAll} 
	\vspace{2mm}
	\resizebox{\textwidth}{!}{
	\begin{tabular}{lccccccc}
		\hline\\[-4mm]
		 & & Beam model & Linkage model& Ref. 1 & Ref. 2 & Ref. 3 & Ref. 4 \\ [1mm]
		\hline\\[-4mm]
		Graphene & \begin{tabular}{@{}c@{}} Z\\A \end{tabular} & \begin{tabular}{@{}c@{}} 0.2331\\0.2405 \end{tabular} & \begin{tabular}{@{}c@{}} 0.2343\\0.2441 \end{tabular} & 0.2385 \citepalias{Kudin2001_01} & 0.225 \citepalias{Lu2009_01} & 0.231 \citepalias{Wei2013_01} & \begin{tabular}{@{}c@{}} 0.2419\\0.2403 \end{tabular} \citepalias{Kumar2020_01} \\[1mm]
		\hline\\[-4mm]
		h-BN & \begin{tabular}{@{}c@{}} Z\\A \end{tabular}  & \begin{tabular}{@{}c@{}} 0.1518\\0.1562 \end{tabular} & \begin{tabular}{@{}c@{}} 0.1602\\0.1627 \end{tabular} & 0.138 \citepalias{Qu2019_01} & \begin{tabular}{@{}c@{}} 0.0897\\0.0929 \end{tabular}  \citepalias{Kumar2020_01} & - & - \\[1mm]
		\hline\\[-4mm]
		Silicene & \begin{tabular}{@{}c@{}} Z\\A \end{tabular} & \begin{tabular}{@{}c@{}} 0.0641 \\0.0644 \end{tabular} & \begin{tabular}{@{}c@{}} 0.0677\\0.0671 \end{tabular} & 6.184 \citepalias{Roman2014_01}& 0.986  \citepalias{Banerjee2016_01} & \begin{tabular}{@{}c@{}}  0.0577 \\ 0.0593 \end{tabular} \citepalias{Kumar2020_01}  & -\\[1mm]
		\hline\\[-4mm]
		Blue phosphorene & \begin{tabular}{@{}c@{}} Z\\A \end{tabular} & \begin{tabular}{@{}c@{}} 0.1142 \\0.1199 \end{tabular} & \begin{tabular}{@{}c@{}} 0.1205\\0.1249 \end{tabular} & 0.1196 \citepalias{Zhang2020_01} & - & - & -\\[1mm]
		\hline

	\end{tabular}}
\end{table}\noindent

\section{Discussion} \label{s:Discussion}
\subsection{Calibration and validation for graphene at small bending deformations}
Fig.~\ref{f:EnergiesVSkappaAllGraphene} shows that graphene is described very accurately at small deformations by both the beam and linkage models, which implies that the proposed methods allow for the accurate determination of the bending stiffness from DFT calculations. This is confirmed by the good agreement with the bending stiffnesses obtained from other approaches in the literature, as Tab.~\ref{t:cMeanStdv} shows. The computational complexity of our proposed approach, however, is generally much lower due to its simplicity and lower number of atoms per unit cell, especially at smaller curvatures.\\
In addition, Tab.~\ref{t:cMeanStdv} shows that the bending stiffness is captured more accurately from the DFT tests if higher atomic configurations (i.e.~bigger units cells) and the linkage model are used. This is due to the following reasons: The localized deformation of graphene is better captured by a discrete linkage model than a continuous beam model. At the same time, the continuous deformation assumption, inherent to the concept of a bending stiffness, is not accurate for lower atomic configurations. The results in Tab.~\ref{t:cMeanStdv} indicate that at least the 12-atom-configuration should be used together with the linkage model to accurately compute the bending stiffness.\\
Moreover, the bending stiffness for the zigzag and armchair directions approach the same value, confirming graphene's known isotropic bending behavior at small deformations. This, however, does not exclude the possibility that zigzag and armchair bending is coupled, which is an aspect that should be investigated in detail in future work. The fact that the accuracy is poor for lower atomic configurations shows that continuum bending models should be used cautiously for deformations exhibiting very large curvatures, even if those deformations are only local.

\subsection{Graphene at moderately large bending deformations}
Fig.~\ref{f:EnergiesVSkappaLarge} shows that the bending stiffnesses calibrated from small deformation theory still capture moderately large bending deformation behavior for graphene very well. This demonstrates that even at moderately large curvatures, the bending energy density of graphene still depends quadratically on the curvature, although now large deformation kinematics should be considered via \hyperref[s:ModifiedBeamModel]{Appendix~A} for the beam model, and Eq.~(\ref{e:SpringEnergyCell}) for the linkage model, which is what is done in Fig.~\ref{f:EnergiesVSkappaLarge}. Based on this, the bending stiffness is still constant, at least as long as the membrane stresses are negligible. Fig.~\ref{f:EnergiesVSkappaLarge} shows that this is the case in our suggested DFT calibration tests, which is why they allow extracting the bending stiffness reliably in the first place. On the other hand, if membrane stresses are significant, coupling between in-plane stretching and bending can be expected to affect bending, leading to an increase in the effective bending resistance. This increase is well understood in structural mechanics, but it remains to be seen if it exactly carries over to 2D lattice materials.

\subsection{Application to other 2D materials}
In addition to graphene, the proposed method has been used to calculate the bending stiffness of h-BN, silicene, and blue phosphorene. The results show good agreement with the works of \citet{Qu2019_01} and \citet{Zhang2020_01} but disagree with other works \citep{Roman2014_01, Banerjee2016_01, Kumar2020_01} (although the disagreement with \citet{Kumar2020_01} is quite small for silicene). The large disagreement with \citet{Roman2014_01} can be attributed to the MD calculations used there, whereas the large disagreement with \citet{Banerjee2016_01} can be attributed to their use of a non-relaxed silicene nanoribbon, which entails a size-effect due to the ribbon boundaries.\\
Despite the corrugation of silicene and blue phosphorene, the bending stiffness of graphene and h-BN is higher, as Tab.~\ref{t:cAll} shows. This implies that (1) the atomic bonds play a more important role in the bending stiffness than the surface geometry, and (2) the atomic bonds of graphene and h-BN provide more resistance to bending than those of silicene and blue phosphorene. The latter point is consistent with the known fact that graphene and h-BN have larger Young's moduli than silicene and blue phosphorene \citep{Falin2017_01,Mortazavi2017_01, Ghaffari2019_01}.\\
Due to the corrugated structure of silicene and blue phosphorene, the bending stiffness in the zigzag direction might be expected to be higher than the bending stiffness in the armchair direction since the second moment of area appears to be larger for bending along the zigzag direction \citep{Woodruff2020_01}. However, the second moment of inertia not only comes from the thickness shown in Fig.~\ref{f:hBNSilicene}. More relevant is the effective thickness of the electron clouds, which can be expected to be similar in both directions. Therefore, no significant difference between the armchair and zigzag bending stiffnesses should appear. This is confirmed by our results in Tab.~\ref{t:cAll}. It again shows the small influence of the surface corrugation.\\
We also note that for very small deformations and energies, computational errors and inherent inaccuracies of DFT might become significant and affect the calculated energies negatively, even though some systematic errors cancel each other when only considering the energy differences in similar unit cells \citep{sholl2011_01}. In the considered infinitesimal regime, however, only results with sufficiently large curvatures have been used to avoid these issues. Moreover, poor choice of DFT convergence parameters can result in wrong bending stiffness values. 

\section{Conclusion} \label{s:Conclusion}
The suggested framework uses relatively small unit cells, is computationally efficient, and is easy to implement in conventional atomistic simulation codes. Therefore, it can be used for the efficient and accurate determination of continuum bending parameters for two-dimensional materials. The framework is very flexible and hence admits many possible extensions: Examples are the incorporation of biaxial bending, larger deformations, membrane-bending coupling, and finite temperatures.
\subsection*{Acknowledgements}
Funding: This work was initially supported by the German Research Foundation through project GSC 111.

\subsection*{Data availability}
All results and representative input files for Quantum Espresso can be provided on request.\\

\appendix
\renewcommand\thefigure{\thesection.\arabic{figure}}  \renewcommand\thetable{\thesection.\arabic{table}}
\renewcommand{\theequation}{\thesection.\arabic{equation}}

\section{Modified beam model for large deformations} \label{s:ModifiedBeamModel} \setcounter{figure}{0} \setcounter{table}{0} \setcounter{equation}{0}
At moderately large deformations, some of the approximations used in the development of the infinitesimal beam model become inaccurate. This section, therefore, presents a more general approach, which is used in the calculation of the energies in Fig.~\ref{f:EnergiesVSkappaLarge}.\\
As a result of bending, the length of the unit cell can change and is now denoted by $4 \ell$. Considering this change in the formulation, the deflection of the beam is still assumed to follow Eq.~(\ref{e:BeamDeformation}).
But the curvature is now taken from the exact expression
\eqb{lll}
\ds \kappa(x) = \frac{w''}{\big(1+(w')^2\big)^{3/2}}\,,
\label{e:kappa1Dfull}
\eqe
where $w'=\dif w/\dif x = 3\,\alpha\,(\ell^2-x^2)/\ell^3 $ and $w''=\dif^2 w/\dif x^2 = -6\, \alpha\, x/\ell^3 $. The bending energy of the unit cell can then be calculated along the curve as
\eqb{lll}
\ds {E_{\mathrm{beam}}} = \int_{0}^{S} \frac{c}{2} \kappa^2 \dif s \,,  
\label{e:BendingEnergyDensitymodified}	
\eqe
where
\eqb{lll} \label{e:S^(k)}
\ds S=\int_{0}^{S} \dif s = \int_0^{\ell} \sqrt{1+(w')^2} \, \dif x  \\
\eqe
is the length of the deflected beam at each load increment.

\section{Equivalent bending stiffness in the rigid linkage model} \label{s:assessment_c} \setcounter{figure}{0} \setcounter{table}{0} \setcounter{equation}{0}
This section derives the equivalent bending stiffness from given spring stiffness $k_\mathrm{s}$ in the rigid linkage model. Fig.~\ref{f:ACCdtheta} shows a segment of the curved atomic surface. Along the armchair direction $a_1 \neq a_2$, while along the zigzag direction $a_1 = a_2$. Assuming that the curvature is constant along $2\bar{a}=a_1+a_2$, the rotation angle of the spring can be obtained from 
\eqb{lll}
\ds 2r \sin{\frac{\gamma_{i}}{2}} = a_{i}  \qquad (i=1,2)\\ [3mm]
\eqe
and
\eqb{lll}
\ds \theta = \frac{\gamma_1}{2}+ \frac{\gamma_2}{2} =\arcsin{\frac{a_{1}}{2r}} + \arcsin{\frac{a_{2}}{2r}}  
	= \arcsin{\bigg(  \frac{a_{1}}{4r^2} \sqrt{4r^2-a_{2}^2}      +   \frac{a_{2}}{4r^2} \sqrt{4r^2-a_{1}^2}       \bigg)\,,}
\eqe
which can then be used in Eq.~(\ref{e:RotationalSpringEnergy0}) to obtain the energy of the spring $U_{\mathrm{s}}$ from the linkage model.
\begin{figure}[h]
	\centering
	\includegraphics[height=55mm]{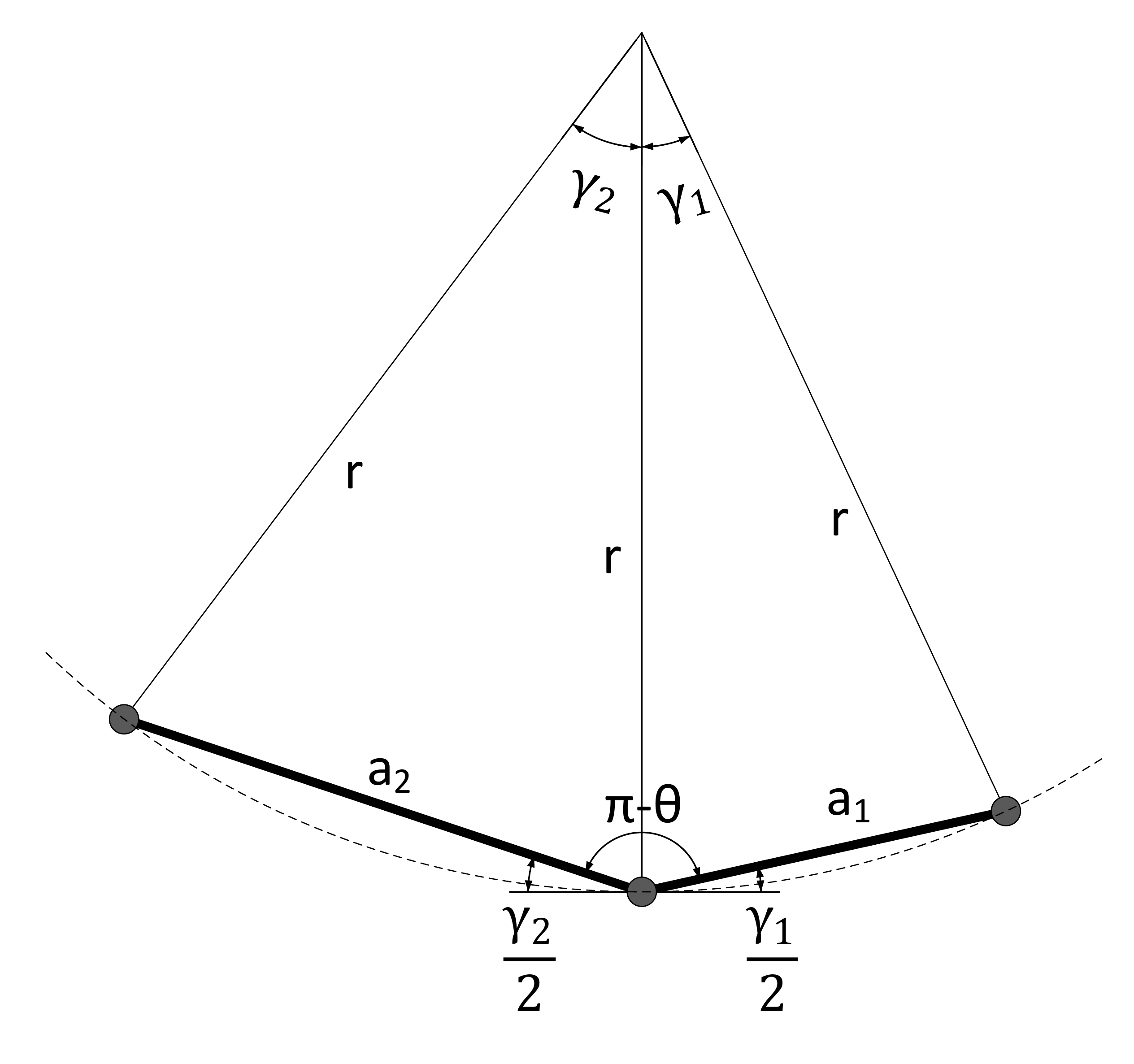}
	\vspace{0mm}
	\caption{An equivalent atomic system with constant curvature. The more general armchair direction with $a_1 \neq a_2$ is shown.}\label{f:ACCdtheta}
\end{figure}
The corresponding energy from the continuum model can be calculated as
\eqb{lll}
\ds U_{\mathrm{c}}=\int_{-\frac{ \gamma_{2}}{2}}^{\frac{ \gamma_{1}}{2}} b\frac{c}{2} \kappa^2 \dif s = b \frac{c}{2} \frac{1}{r^2} r \bigg(\frac{\gamma_{1}+\gamma_{2}}{2}\bigg) = \frac{b\,c}{2r} \, \theta\,. \\[3mm]
\eqe
Setting $U_{\mathrm{c}}=U_{\mathrm{s}}$ gives 
\eqb{lll} \label{e:Acc_c_K}
\ds c=k_{\mathrm{s}} r \arcsin{ \bigg(  \frac{a_{1}}{4r^2} \sqrt{4r^2-a_{2}^2}      +   \frac{a_{2}}{4r^2} \sqrt{4r^2-a_{1}^2}       \bigg). } \\ [3mm]
\eqe
It can be seen that the bending stiffness generally depends on the bending radius. For small deformations, however, $a_1/r\,,\, a_2/r \ll 1 $ and $\arcsin{(\bar a/r)} \approx \bar a/r$ such that Eq.~(\ref{e:Acc_c_K}) tends to 
\eqb{lll} \label{e:est_c_K}
\ds c=k_{\mathrm{s}} 
\bar{a}\,. \\ [3mm]
\eqe
Even for large curvatures up to $\bar{a}/r<0.5$, the difference between (\ref{e:Acc_c_K}) and (\ref{e:est_c_K}) is less than 1.5\%. 

\section{Average curvature, angle, and energy density} \label{s:Supplementary} \setcounter{figure}{0} \setcounter{table}{0} \setcounter{equation}{0}
This section presents the definition of the three average quantities used in Figs.~\ref{f:EnergiesVSkappaAllGraphene}-\ref{f:BNSiliceneEPlot}. The average curvature is a useful measure to compare different atomic configurations using the beam model, and it is defined at each load increment as
\eqb{lll} \label{e:Averagekappa}
\ds \bar{\kappa} := - \frac{1}{S}\int_0^{S} \kappa(s)\,\dif s \approx - \frac{1}{L}\int_0^{L} \kappa(x)\,\dif x = \frac{3\alpha}{L^2}\,.
\eqe
The curvature in $0 \leq x \leq L$ is negative for the deformations of Fig.~\ref{f:AllModesBC}, hence the negative sign is added to make $\bar\kappa$ positive. For the rigid linkage model, the average angle change 
\eqb{lll} \label{e:AngleChange}
\ds \bar{\theta} := \frac{1}{n}\sum_{i=1}^{n}  \theta_{i} 
\eqe
is used for the comparison of different atomic configurations. Under the small deformation assumption, follows
\eqb{lll} \label{e:AngleChangeSmall}
\ds \bar{\theta} = \chi \frac{w_0}{\bar a}\, , \\ [3mm]
\eqe
where the values of $\chi$ are tabulated in Tab.~\ref{t:Theta_constant}. Multiplying (\ref{e:Averagekappa}) by $1 = \bar a \bar\theta/(\chi w_0)$, which follows from (\ref{e:AngleChangeSmall}), gives the relation between the two average measures, 
\eqb{lll} \label{e:kappaVStheta}
\ds \bar\kappa =  \frac{ 3 \, \alpha\, \bar a}{\chi\, w_0\,  L^2}    \, \bar\theta \,,
\eqe
where $\alpha/w_0$ is tabulated in Tab.~\ref{t:alfas}.\\

\begin{table}[h] 
	\centering
	\small
	\caption{Values of $\chi$ in the rigid linkage model for different atomic configurations} \label{t:Theta_constant} 
	\vspace{-2mm}
	\begin{tabular}{lclc}
		\hline \\ [-3mm]
		Zigzag direction & $\chi$ & Armchair direction & $\chi$ \\ [1mm]
		\hline \\ [-3mm]
		4-atom-config. & 1 & 4-atom-config. & 3/2  \\ [1mm]
		\hline \\ [-3mm]
		6-atom-config. & 2/3 &   & \\ [1mm] 
		\hline \\ [-3mm]
		8-atom-config. & 1/3 & 8-atom-config. & 0.3804 \\ [1mm]
		\hline \\ [-3mm]
		10-atom-config. & 6/25 & &\\ [1mm]
		\hline \\ [-3mm]
		12-atom-config. & 3/19 & 12-atom-config. & 0.1602 \\ [1mm]
		\hline \\ [-3mm]
		14-atom-config. & 6/49 & & \\ [1mm]
		\hline \\ [-3mm]
		16-atom-config. & 1/11 & 16-atom-config. & 0.09 \\ [1mm]
		\hline	
	\end{tabular}
\end{table}\noindent
The third quantity, the average energy density, is defined by
\eqb{lll} \label{e:AverageEnergyDensity}
\ds \bar{W_\bullet} := \frac{E_\bullet}{4Lb}\,, \\ [3mm]
\eqe
where $E_\bullet$ collectively denotes various cases: the bending energies of the beam (Eqs.~(\ref{e:BeamEnergyCell}) and (\ref{e:BendingEnergyDensitymodified})) or linkage model (Eqs.~(\ref{e:SpringEnergyCellSimplified}) and (\ref{e:SpringEnergyCell})), or the membrane energy from the linkage model (Eq.~(\ref{e:TotalMembraneEnergy})).

\section{Calculation of membrane energies at large deformations} \label{s:MembraneEnergy} \setcounter{figure}{0} \setcounter{table}{0} \setcounter{equation}{0}
This section presents the calculation of the membrane strain energy, which is essential for the assessment of large deformations. Since the linkage kinematics in Fig.~\ref{f:SM} is not approximate, the linkage model is used for the calculation.\\
Allowing now for stretchable links, the stretch of the links and the stretch in lateral direction follow as
\eqb{lll} \label{e:Lambda1Lambda2}
\ds (\lambda_1)_i=\frac{d_i}{(d_0)_i} \qquad  \qquad
\ds \lambda_2=\frac{b}{b_0} \,,
\eqe
where $d_i$, $(d_0)_i$, $b$, and $b_0$ are the current and the initial length of link $i$, and the current and the initial unit cell width, respectively. They follow from the DFT data. The total membrane energy is the sum of the membrane energies of all links and is then given by
\eqb{lll}\label{e:TotalMembraneEnergy}
\ds \Psi_{\mathrm{linkage}} =b \sum_{i=1}^{N_{\mathrm{links}}}  d_i W_{\mathrm{m}} \big((\lambda_1)_i,\lambda_2\big)  \,,
\eqe
where $N_{\mathrm{links}}$ is the number of links and $W_{\mathrm{m}}$ is the membrane energy density from a suitable material model, such as the one for graphene presented in \citep{Shirazian2018_01}. 

\section{Bending stiffness values of h-BN, silicene, and blue phosphorene for different atomic configurations} \label{s:extra_tab_fig} \setcounter{figure}{0} \setcounter{table}{0} \setcounter{equation}{0}
The bending stiffnesses of h-BN, silicene, and blue phosphorene have been calculated for the 8- and 12-atom configurations along the zigzag, and 12- and 16-atom configurations along the armchair direction. The values for the beam and linkage model are given in Tab.~\ref{t:BendingStiffnessOthers}.
\begin{table}[h] 
	\centering
	\caption{The calibrated bending stiffness $c~~[\mathrm{nN.nm}]$ for different atomic configurations of h-BN, silicene, and blue phosphorene.} \label{t:BendingStiffnessOthers} 
	\vspace{2mm}
	\begin{tabular}{lccc}
		\hline \\ [-4mm]
		 & h-BN & Silicene & Blue phosphorene \\ [1mm]
		\hline \\ [-4mm]
		8-atom-config. zigzag | beam model & 0.1448 & 0.0698 & 0.1734 \\ [1mm]
		\hline \\ [-4mm]
		8-atom-config. zigzag | linkage model & 0.1629 & 0.0785  & 0.1950 \\ [1mm]
		\hline \\ [-4mm]
		12-atom-config. zigzag | beam model & 0.1518 & 0.0641 & 0.1142 \\ [1mm]
		\hline \\ [-4mm]
		12-atom-config. zigzag | linkage model & 0.1602 & 0.0677 & 0.1205 \\ [1mm]
		\hline \\ [-4mm]
		12-atom-config. armchair | beam model & 0.1370 & 0.0641 & 0.1203 \\ [1mm]
		\hline \\ [-4mm]
		12-atom-config. armchair | linkage model & 0.1480 & 0.0692 & 0.1300 \\ [1mm]
		\hline \\ [-4mm]
		16-atom-config. armchair | beam model & 0.1562 & 0.0644 & 0.1199 \\ [1mm]
		\hline \\ [-4mm]
		16-atom-config. armchair | linkage model & 0.1627 & 0.0671 & 0.1249 \\ [1mm]
		\hline \\ [-4mm]
	\end{tabular}
\end{table}\noindent
\bibliographystyle{apalike}
\bibliography{Manuscript.bib}

\begin{thebibliography}{}

\bibitem[Ahmadpoor et~al., 2017]{Ahmadpoor2017_01}
Ahmadpoor, F., Wang, P., Huang, R., and Sharma, P. (2017).
\newblock Thermal fluctuations and effective bending stiffness of elastic thin
  sheets and graphene: A nonlinear analysis.
\newblock {\em J. Mech. Phys. Solids}, {\bf 107}:294--319.

\bibitem[Akinwande et~al., 2017]{Akinwande2017_01}
Akinwande, D., Brennan, C.~J., Bunch, J.~S., Egberts, P., Felts, J.~R., Gao,
  H., Huang, R., Kim, J.-S., Li, T., Li, Y., Liechti, K.~M., Lu, N., Park,
  H.~S., Reed, E.~J., Wang, P., Yakobson, B.~I., Zhang, T., Zhang, Y.-W., Zhou,
  Y., and Zhu, Y. (2017).
\newblock A review on mechanics and mechanical properties of 2d
  materials—graphene and beyond.
\newblock {\em Extreme Mech. Lett.}, {\bf 13}:42 -- 77.

\bibitem[Arroyo and Belytschko, 2004]{Arroyo2004_01}
Arroyo, M. and Belytschko, T. (2004).
\newblock Finite crystal elasticity of carbon nanotubes based on the
  exponential {C}auchy-{B}orn rule.
\newblock {\em Phys. Rev. B}, {\bf 69}{\normalfont (11)}:115415.

\bibitem[Banerjee and Suryanarayana, 2016]{Banerjee2016_01}
Banerjee, A.~S. and Suryanarayana, P. (2016).
\newblock Cyclic density functional theory: A route to the first principles
  simulation of bending in nanostructures.
\newblock {\em J. Mech. Phys. Solids}, {\bf 96}:605--631.

\bibitem[Berinskii et~al., 2014]{Berinskii2014_01}
Berinskii, I.~E., Krivtsov, A.~M., and Kudarova, A.~M. (2014).
\newblock Bending stiffness of a graphene sheet.
\newblock {\em Phys. Mesomech.}, {\bf 17}(4):356--364.

\bibitem[Bl\"ochl, 1994]{Blochl1994_01}
Bl\"ochl, P.~E. (1994).
\newblock Projector augmented-wave method.
\newblock {\em Phys. Rev. B}, {\bf 50}{\normalfont (24)}:17953--17979.

\bibitem[Burmistrov et~al., 2018]{Burmistrov2018_01}
Burmistrov, I.~S., Gornyi, I.~V., Kachorovskii, V.~Y., Katsnelson, M.~I., Los,
  J.~H., and Mirlin, A.~D. (2018).
\newblock Stress-controlled {P}oisson ratio of a crystalline membrane:
  Application to graphene.
\newblock {\em Phys. Rev. B}, {\bf 97}{\normalfont (12)}:125402.

\bibitem[Chandler and Vella, 2020]{Chandler2020_01}
Chandler, T. and Vella, D. (2020).
\newblock Indentation of suspended two-dimensional solids: The signatures of
  geometrical and material nonlinearity.
\newblock {\em J. Mech. Phys. Solids}, {\bf 144}:104109.

\bibitem[Chen et~al., 2017]{Chen2017_01}
Chen, J., Wang, B., and Hu, Y. (2017).
\newblock An existence criterion for low-dimensional materials.
\newblock {\em J. Mech. Phys. Solids}, {\bf 107}:451--468.

\bibitem[Chen et~al., 2020]{Chen2020_01}
Chen, Z., Wang, H., and Li, Z. (2020).
\newblock First-principles study of two dimensional {C3N} and its derivatives.
\newblock {\em RSC Adv.}, {\bf 10}{\normalfont (55)}:33469--33474.

\bibitem[Davini et~al., 2017]{Davini2017_01}
Davini, C., Favata, A., and Paroni, R. (2017).
\newblock The {G}aussian stiffness of graphene deduced from a continuum model
  based on molecular dynamics potentials.
\newblock {\em J. Mech. Phys. Solids}, {\bf 104}:96 -- 114.

\bibitem[Doedens et~al., 2017]{Doedens2017_01}
Doedens, R.~J., Eaton, P.~E., and Fleischer, E.~B. (2017).
\newblock The bent bonds of cubane.
\newblock {\em Eur. J. Org. Chem.}, {\bf 2017}(18):2627--2630.

\bibitem[Falin et~al., 2017]{Falin2017_01}
Falin, A., Cai, Q., Santos, E. J.~G., Scullion, D., Qian, D., Zhang, R., Yang,
  Z., Huang, S., Watanabe, K., Taniguchi, T., Barnett, M.~R., Chen, Y., Ruoff,
  R.~S., and Li, L.~H. (2017).
\newblock Mechanical properties of atomically thin boron nitride and the role
  of interlayer interactions.
\newblock {\em Nat. Commun.}, {\bf 8}(1):15815.

\bibitem[Gao and Huang, 2014]{Gao2014_01}
Gao, W. and Huang, R. (2014).
\newblock Thermomechanics of monolayer graphene: Rippling, thermal expansion
  and elasticity.
\newblock {\em J. Mech. Phys. Solids}, {\bf 66}:42--58.

\bibitem[Garc{\'i}a~de Abajo, 2014]{GarciadeAbajo2014_01}
Garc{\'i}a~de Abajo, F.~J. (2014).
\newblock Graphene plasmonics: Challenges and opportunities.
\newblock {\em ACS Photonics}, {\bf 1}(3):135--152.

\bibitem[Ghaffari et~al., 2018]{Ghaffari2018_01}
Ghaffari, R., Duong, T.~X., and Sauer, R.~A. (2018).
\newblock A new shell formulation for graphene structures based on existing
  ab-initio data.
\newblock {\em Int. J. Solids Struct.}, {\bf 135}:37--60.

\bibitem[Ghaffari et~al., 2019]{Ghaffari2019_01}
Ghaffari, R., Shirazian, F., Hu, M., and Sauer, R.~A. (2019).
\newblock A nonlinear hyperelasticity model for single layer blue phosphorus
  based on \textit{ab initio} calculations.
\newblock {\em Proc. R. Soc. A: Math. Phys. Eng. Sci.}, {\bf
  475}(2229):20190149.

\bibitem[Ghosh et~al., 2019]{Ghosh2019_01}
Ghosh, S., Banerjee, A.~S., and Suryanarayana, P. (2019).
\newblock Symmetry-adapted real-space density functional theory for cylindrical
  geometries: Application to large group-iv nanotubes.
\newblock {\em Phys. Rev. B}, {\bf 100}{\normalfont (12)}:125143.

\bibitem[Goerbig, 2011]{Goerbig2011_01}
Goerbig, M.~O. (2011).
\newblock Electronic properties of graphene in a strong magnetic field.
\newblock {\em Rev. Mod. Phys.}, {\bf 83}{\normalfont (4)}:1193--1243.

\bibitem[Grigorenko et~al., 2012]{Grigorenko2012_01}
Grigorenko, A.~N., Polini, M., and Novoselov, K.~S. (2012).
\newblock Graphene plasmonics.
\newblock {\em Nat. Photonics}, {\bf 6}(11):749--758.

\bibitem[Grima et~al., 2015]{Grima2015_01}
Grima, J.~N., Winczewski, S., Mizzi, L., Grech, M.~C., Cauchi, R., Gatt, R.,
  Attard, D., Wojciechowski, K.~W., and Rybicki, J. (2015).
\newblock Tailoring graphene to achieve negative {P}oisson's ratio properties.
\newblock {\em Adv. Mater.}, {\bf 27}(8):1455--1459.

\bibitem[Gui et~al., 2008]{Gui2008_01}
Gui, G., Li, J., and Zhong, J. (2008).
\newblock Band structure engineering of graphene by strain: First-principles
  calculations.
\newblock {\em Phys. Rev. B}, {\bf 78}{\normalfont (7)}:075435.

\bibitem[Han et~al., 2020]{Han2020_01}
Han, E., Yu, J., Annevelink, E., Son, J., Kang, D.~A., Watanabe, K., Taniguchi,
  T., Ertekin, E., Huang, P.~Y., and van~der Zande, A.~M. (2020).
\newblock Ultrasoft slip-mediated bending in few-layer graphene.
\newblock {\em Nat. Mater.}, {\bf 19}(3):305--309.

\bibitem[Hinchet et~al., 2018]{Hinchet2018_01}
Hinchet, R., Khan, U., Falconi, C., and Kim, S.-W. (2018).
\newblock Piezoelectric properties in two-dimensional materials: {S}imulations
  and experiments.
\newblock {\em Mater. Today}, {\bf 21}(6):611--630.

\bibitem[Jiang et~al., 2016]{Jiang2016_01}
Jiang, J.-W., Chang, T., Guo, X., and Park, H.~S. (2016).
\newblock Intrinsic negative {Poisson’s} ratio for single-layer graphene.
\newblock {\em Nano Lett.}, {\bf 16}(8):5286--5290.

\bibitem[Jiang and Park, 2014]{Jiang2014_01}
Jiang, J.-W. and Park, H.~S. (2014).
\newblock Negative {Poisson's} ratio in single-layer black phosphorus.
\newblock {\em Nat. Commun.}, {\bf 5}(1):4727.

\bibitem[Jin et~al., 2020]{Jin2020_01}
Jin, W., Sun, W., Kuang, X., Lu, C., and Kou, L. (2020).
\newblock Negative {Poisson} ratio in two-dimensional tungsten nitride:
  Synergistic effect from electronic and structural properties.
\newblock {\em J. Phys. Chem. Lett.}, {\bf 11}(22):9643--9648.

\bibitem[Kit et~al., 2011]{Kit2011_01}
Kit, O.~O., Pastewka, L., and Koskinen, P. (2011).
\newblock Revised periodic boundary conditions: Fundamentals, electrostatics,
  and the tight-binding approximation.
\newblock {\em Phys. Rev. B}, {\bf 84}{\normalfont (15)}:155431.

\bibitem[Koskinen and Kit, 2010a]{Koskinen2010_01}
Koskinen, P. and Kit, O.~O. (2010a).
\newblock Approximate modeling of spherical membranes.
\newblock {\em Phys. Rev. B}, {\bf 82}{\normalfont (23)}:235420.

\bibitem[Koskinen and Kit, 2010b]{Koskinen2010_02}
Koskinen, P. and Kit, O.~O. (2010b).
\newblock Efficient approach for simulating distorted materials.
\newblock {\em Phys. Rev. Lett.}, {\bf 105}{\normalfont (10)}:106401.

\bibitem[Kudin et~al., 2001]{Kudin2001_01}
Kudin, K.~N., Scuseria, G.~E., and Yakobson, B.~I. (2001).
\newblock {${\mathrm{C}}_{2}\mathrm{F},$} {BN}, and {C} nanoshell elasticity
  from ab initio computations.
\newblock {\em Phys. Rev. B}, {\bf 64}{\normalfont (23)}:235406.

\bibitem[Kumar and Parks, 2015]{Kumar2015_01}
Kumar, S. and Parks, D.~M. (2015).
\newblock On the hyperelastic softening and elastic instabilities in graphene.
\newblock {\em Proc. R. Soc. A: Math. Phys. Eng. Sci.}, {\bf
  471}(2173):20140567.

\bibitem[Kumar and Suryanarayana, 2020]{Kumar2020_01}
Kumar, S. and Suryanarayana, P. (2020).
\newblock Bending moduli for forty-four select atomic monolayers from first
  principles.
\newblock {\em Nanotechnology}, {\bf 31}(43):43LT01.

\bibitem[Lindahl et~al., 2012]{Lindahl2012_01}
Lindahl, N., Midtvedt, D., Svensson, J., Nerushev, O.~A., Lindvall, N.,
  Isacsson, A., and Campbell, E. E.~B. (2012).
\newblock Determination of the bending rigidity of graphene via electrostatic
  actuation of buckled membranes.
\newblock {\em Nano Lett.}, {\bf 12}(7):3526--3531.

\bibitem[Liu et~al., 2014]{Liu2014_01}
Liu, X., Pan, D., Hong, Y., and Guo, W. (2014).
\newblock Bending {P}oisson effect in two-dimensional crystals.
\newblock {\em Phys. Rev. Lett.}, {\bf 112}{\normalfont (20)}:205502.

\bibitem[Lu et~al., 2009]{Lu2009_01}
Lu, Q., Arroyo, M., and Huang, R. (2009).
\newblock Elastic bending modulus of monolayer graphene.
\newblock {\em J. Phys. D: Appl. Phys.}, {\bf 42}(10):102002.

\bibitem[Mir et~al., 2020]{Mir2020_01}
Mir, S.~H., Yadav, V.~K., and Singh, J.~K. (2020).
\newblock Recent advances in the carrier mobility of two-dimensional materials:
  A theoretical perspective.
\newblock {\em ACS Omega}, {\bf 5}(24):14203--14211.

\bibitem[Mokhalingam et~al., 2020]{Mokhalingam2020_01}
Mokhalingam, A., Ghaffari, R., Sauer, R.~A., and Gupta, S.~S. (2020).
\newblock Comparing quantum, molecular and continuum models for graphene at
  large deformations.
\newblock {\em Carbon}, {\bf 159}:478--494.

\bibitem[Momma and Izumi, 2011]{Momma2011_01}
Momma, K. and Izumi, F. (2011).
\newblock Vesta 3 for three-dimensional visualization of crystal, volumetric
  and morphology data.
\newblock {\em J. Appl. Crystallogr.}, {\bf 44}(6):1272--1276.

\bibitem[Monkhorst and Pack, 1976]{Monkhorst1976_01}
Monkhorst, H.~J. and Pack, J.~D. (1976).
\newblock Special points for {B}rillouin-zone integrations.
\newblock {\em Phys. Rev. B}, {\bf 13}{\normalfont (12)}:5188--5192.

\bibitem[Mortazavi et~al., 2021]{Mortazavi2021_01}
Mortazavi, B., Javvaji, B., Shojaei, F., Rabczuk, T., Shapeev, A.~V., and
  Zhuang, X. (2021).
\newblock Exceptional piezoelectricity, high thermal conductivity and stiffness
  and promising photocatalysis in two-dimensional {MoSi2N4} family confirmed by
  first-principles.
\newblock {\em Nano Energy}, {\bf 82}:105716.

\bibitem[Mortazavi et~al., 2017]{Mortazavi2017_01}
Mortazavi, B., Rahaman, O., Makaremi, M., Dianat, A., Cuniberti, G., and
  Rabczuk, T. (2017).
\newblock First-principles investigation of mechanical properties of silicene,
  germanene and stanene.
\newblock {\em Phys. E: Low-dimens. Syst. Nanostructures}, {\bf 87}:228--232.

\bibitem[Muñoz et~al., 2010]{Munoz2010_01}
Muñoz, E., Singh, A.~K., Ribas, M.~A., Penev, E.~S., and Yakobson, B.~I.
  (2010).
\newblock The ultimate diamond slab: Graphane versus graphene.
\newblock {\em Diam. Relat. Mater.}, {\bf 19}(5):368--373.
\newblock Proceedings of Diamond 2009, The 20th European Conference on Diamond,
  Diamond-Like Materials, Carbon Nanotubes and Nitrides, Part 1.

\bibitem[Nikiforov et~al., 2014]{Nikiforov2014_01}
Nikiforov, I., Dontsova, E., James, R.~D., and Dumitric\u{a}, T. (2014).
\newblock Tight-binding theory of graphene bending.
\newblock {\em Phys. Rev. B}, {\bf 89}{\normalfont (15)}:155437.

\bibitem[Novoselov et~al., 2005]{Novoselov2005_01}
Novoselov, K.~S., Jiang, D., Schedin, F., Booth, T.~J., Khotkevich, V.~V.,
  Morozov, S.~V., and Geim, A.~K. (2005).
\newblock Two-dimensional atomic crystals.
\newblock {\em Proc. Natl. Acad. Sci.}, {\bf 102}(30):10451--10453.

\bibitem[{P. Giannozzi et. al.}, 2009]{Giannozzi2009_01}
{P. Giannozzi et. al.} (2009).
\newblock {QUANTUM} {ESPRESSO}: a modular and open-source software project for
  quantum simulations of materials.
\newblock {\em J. Phys. Condens. Matter}, {\bf 21}(39):395502.

\bibitem[{P. Giannozzi et. al.}, 2017]{Giannozzi2017_01}
{P. Giannozzi et. al.} (2017).
\newblock Advanced capabilities for materials modelling with quantum
  {ESPRESSO}.
\newblock {\em J. Phys. Condens. Matter}, {\bf 29}(46):465901.

\bibitem[Perdew et~al., 1996]{Perdew1996_01}
Perdew, J.~P., Burke, K., and Ernzerhof, M. (1996).
\newblock Generalized gradient approximation made simple.
\newblock {\em Phys. Rev. Lett.}, {\bf 77}{\normalfont (18)}:3865--3868.

\bibitem[Pini et~al., 2016]{Pini2016_01}
Pini, V., Ruz, J.~J., Kosaka, P.~M., Malvar, O., Calleja, M., and Tamayo, J.
  (2016).
\newblock How two-dimensional bending can extraordinarily stiffen thin sheets.
\newblock {\em Sci. Rep.}, {\bf 6}(1):29627.

\bibitem[Qu et~al., 2019]{Qu2019_01}
Qu, W., Bagchi, S., Chen, X., Chew, H.~B., and Ke, C. (2019).
\newblock Bending and interlayer shear moduli of ultrathin boron nitride
  nanosheet.
\newblock {\em J. Phys. D: Appl. Phys.}, {\bf 52}(46):465301.

\bibitem[Roman and Cranford, 2014]{Roman2014_01}
Roman, R.~E. and Cranford, S.~W. (2014).
\newblock Mechanical properties of silicene.
\newblock {\em Comput. Mater. Sci.}, {\bf 82}:50--55.

\bibitem[Scarpa et~al., 2010]{Scarpa2010_01}
Scarpa, F., Adhikari, S., Gil, A.~J., and Remillat, C. (2010).
\newblock The bending of single layer graphene sheets: the lattice versus
  continuum approach.
\newblock {\em Nanotechnology}, {\bf 21}(12):125702.

\bibitem[Shirazian et~al., 2018]{Shirazian2018_01}
Shirazian, F., Ghaffari, R., Hu, M., and Sauer, R.~A. (2018).
\newblock Hyperelastic material modeling of graphene based on density
  functional calculations.
\newblock {\em PAMM}, {\bf18}(1):e201800419.

\bibitem[Sholl and Steckel, 2011]{sholl2011_01}
Sholl, D. and Steckel, J.~A. (2011).
\newblock {\em Density functional theory: a practical introduction}.
\newblock John Wiley \& Sons.

\bibitem[Sohier et~al., 2017]{Sohier2017_01}
Sohier, T., Calandra, M., and Mauri, F. (2017).
\newblock Density functional perturbation theory for gated two-dimensional
  heterostructures: Theoretical developments and application to flexural
  phonons in graphene.
\newblock {\em Phys. Rev. B}, {\bf 96}{\normalfont (7)}:075448.

\bibitem[Sorokin and Yakobson, 2021]{Sorokin2021_01}
Sorokin, P.~B. and Yakobson, B.~I. (2021).
\newblock Two-dimensional diamond---diamane: Current state and further
  prospects.
\newblock {\em Nano Lett.}, {\bf 21}(13):5475--5484.

\bibitem[Tong et~al., 2021]{Tong2021_01}
Tong, Z., Dumitric{\u{a}}, T., and Frauenheim, T. (2021).
\newblock Ultralow thermal conductivity in two-dimensional {MoO3}.
\newblock {\em Nano Lett.}, {\bf 21}(10):4351--4356.

\bibitem[Wang et~al., 2017a]{Wang2017_02}
Wang, H., Li, X., Li, P., and Yang, J. (2017a).
\newblock $\delta$-{P}hosphorene: a two dimensional material with a highly
  negative {P}oisson{'}s ratio.
\newblock {\em Nanoscale}, {\bf 9}{\normalfont (2)}:850--855.

\bibitem[Wang et~al., 2017b]{Wang2017_01}
Wang, H., Li, X., Liu, Z., and Yang, J. (2017b).
\newblock $\psi$-{P}hosphorene: a new allotrope of phosphorene.
\newblock {\em Phys. Chem. Chem. Phys.}, {\bf 19}{\normalfont (3)}:2402--2408.

\bibitem[Wang et~al., 2012]{Wang2012_01}
Wang, Q.~H., Kalantar-Zadeh, K., Kis, A., Coleman, J.~N., and Strano, M.~S.
  (2012).
\newblock Electronics and optoelectronics of two-dimensional transition metal
  dichalcogenides.
\newblock {\em Nat. Nanotechnol.}, {\bf 7}(11):699--712.

\bibitem[Wei et~al., 2013]{Wei2013_01}
Wei, Y., Wang, B., Wu, J., Yang, R., and Dunn, M.~L. (2013).
\newblock Bending rigidity and {Gaussian} bending stiffness of single-layered
  graphene.
\newblock {\em Nano Lett.}, {\bf 13}(1):26--30.

\bibitem[Wiberg, 1996]{Wiberg1996_01}
Wiberg, K.~B. (1996).
\newblock Bent bonds in organic compounds.
\newblock {\em Acc. Chem. Res.}, {\bf 29}(5):229--234.

\bibitem[Woodruff and Filipov, 2020]{Woodruff2020_01}
Woodruff, S.~R. and Filipov, E.~T. (2020).
\newblock Curved creases redistribute global bending stiffness in corrugations:
  theory and experimentation.
\newblock {\em Meccanica}, {\bf 56}(6):1613--1634.

\bibitem[Yang et~al., 2017]{Yang2017_01}
Yang, S., Li, W., Ye, C., Wang, G., Tian, H., Zhu, C., He, P., Ding, G., Xie,
  X., Liu, Y., Lifshitz, Y., Lee, S.-T., Kang, Z., and Jiang, M. (2017).
\newblock {C3N}—a {2D} crystalline, hole-free, tunable-narrow-bandgap
  semiconductor with ferromagnetic properties.
\newblock {\em Adv. Mater.}, {\bf 29}(16):1605625.

\bibitem[Yu et~al., 2016]{Yu2016_01}
Yu, L., Ruzsinszky, A., and Perdew, J.~P. (2016).
\newblock Bending two-dimensional materials to control charge localization and
  fermi-level shift.
\newblock {\em Nano Lett.}, {\bf 16}(4):2444--2449.

\bibitem[Zhang et~al., 2011]{Zhang2011_01}
Zhang, D.-B., Akatyeva, E., and Dumitric\u{a}, T. (2011).
\newblock Bending ultrathin graphene at the margins of continuum mechanics.
\newblock {\em Phys. Rev. Lett.}, {\bf 106}{\normalfont (25)}:255503.

\bibitem[Zhang and Jiang, 2015]{Zhang2015_01}
Zhang, H.-Y. and Jiang, J.-W. (2015).
\newblock Elastic bending modulus for single-layer black phosphorus.
\newblock {\em J. Phys. D: Appl. Phys.}, {\bf 48}(45):455305.

\bibitem[Zhang et~al., 2020]{Zhang2020_01}
Zhang, T., Di, X., Chen, G., and Zhu, L. (2020).
\newblock Parameterization of a {COMPASS} force field for single layer blue
  phosphorene.
\newblock {\em Nanotechnology}, {\bf 31}(14):145702.

\bibitem[Zhang et~al., 2017]{Zhang2017_01}
Zhang, Z., Penev, E.~S., and Yakobson, B.~I. (2017).
\newblock Two-dimensional boron: structures{,} properties and applications.
\newblock {\em Chem. Soc. Rev.}, {\bf 46}{\normalfont (22)}:6746--6763.

\bibitem[Zhao et~al., 2015]{Zhao2015_01}
Zhao, J., Deng, Q., Ly, T.~H., Han, G.~H., Sandeep, G., and R{\"u}mmeli, M.~H.
  (2015).
\newblock Two-dimensional membrane as elastic shell with proof on the folds
  revealed by three-dimensional atomic mapping.
\newblock {\em Nat. Commun.}, {\bf 6}(1):8935.

\end{thebibliography}
\end{document}